\documentclass{pasj00}
\draft

\begin{document}
\SetRunningHead{Fujita et al.}{Running Head}
\Received{2000/01/01}
\Accepted{2000/01/01}

\title{Chandra Observations of A2670 and A2107:
A Comet Galaxy and cDs with Large Peculiar Velocities}

 \author{%
   Yutaka \textsc{Fujita}\altaffilmark{1}
   Craig L. \textsc{Sarazin}\altaffilmark{2}
   and
   Gregory R. \textsc{Sivakoff}\altaffilmark{2}}
 \altaffiltext{1}{Department of Earth and Space Science, 
Graduate School of Science,\\
Osaka University, 
Toyonaka, Osaka 560-0043}
 \email{fujita@vega.ess.sci.osaka-u.ac.jp}
 \altaffiltext{2}{Department of Astronomy, University of
Virginia, \\
P. O. Box 3818, Charlottesville, VA 22903-0818, USA}


\maketitle

\begin{abstract}
 We present an analysis of Chandra observations of the galaxy
 clusters A2670 and A2107. Their cD galaxies have large peculiar
 velocities ($>200~\rm\: km\: s^{-1}$) and thus the clusters appear to
 be undergoing mergers.  In A2670, we find a comet-like structure around
 one of the brightest galaxies. At the leading edge of the structure,
 there is a cold front. The mass of the X-ray gas in the comet-like
 structure suggests that the galaxy was in a small cluster or group, and
 its intracluster medium (ICM) is being stripped by ram-pressure. The
 regions of cool interstellar medium (ISM) of the cD galaxies in A2670
 and A2107 are very compact. This is similar to the brightest galaxies
 in the Coma cluster, which is also a merging cluster. In each galaxy,
 the short cooling time of the ISM requires a heating source; the
 compact nature of the ISM makes it unlikely that the heating source is
 a central active galactic nucleus (AGN).
\end{abstract}

\section{Introduction}

It is generally believed that dark matter constitutes a large fraction
of the mass in the Universe. Among various theories of dark matter, cold
dark matter (CDM) theory provides a remarkably successful description of
large-scale structure formation and is in good agreement with a large
variety of observational data. This model predicts that small objects
are the first to form and that these then amalgamate into progressively
larger systems. In this model, clusters of galaxies are considered to be
the objects that have recently formed via mergers of subclusters.  A
cluster merger is one of the most spectacular events in the Universe. In
a major merger, subclusters collide at velocities of $\gtrsim 1000\rm\;
km\; s^{-1}$ and release gravitational energies of $\sim 10^{64}$~erg.

A giant elliptical galaxy (cD galaxy) is often located at the center of
a cluster. In general, the peculiar velocities of the cD galaxies are
much smaller ($\lesssim 200\:\rm km\: s^{-1}$; \authorcite{oeg01}
\yearcite{oeg01}) than the velocity dispersions of other galaxies in the
clusters ($\sim 1000\:\rm km\: s^{-1}$). This means that the cD galaxies
are nearly at rest in the cluster potential wells. However, the cD
galaxies in some clusters have large peculiar velocities ($\gtrsim
200\:\rm km\: s^{-1}$).  Since a cluster merger may disturb a cD galaxy
from its resting place at the bottom of the potential, the large
peculiar velocity of the cD galaxy is a good indicator of mergers.

In this paper, we present the results of Chandra observations of two
clusters containing rapidly moving cD galaxies in order to study the
effects of cluster mergers on the cluster galaxies (including the cD
galaxies) and the nature of substructures within clusters.  Recently,
Oegerle and Hill (\yearcite{oeg01}) studied the redshifts of cluster
galaxies and their distributions intensively. They observed redshifts of
$50$--$300$ galaxies per cluster and studied the peculiar velocities of
cD galaxies in the clusters.  For several clusters, they compiled
published data. Among 25 clusters, they found 4 clusters for which the
redshifts of the cD galaxies are significantly different from the
cluster means within $1.5\; h_{75}^{-1}$~Mpc of the cDs, where the
Hubble constant is $H_0=75\; h_{75}\;\rm km\; s^{-1}\; Mpc^{-1}$. The
clusters are A2052, A2107, A2199, and A2670. The peculiar velocities of
the cD galaxies are $\sim 250$--$400\;\rm km\; s^{-1}$. Among them,
A2052 and A2199 have already been observed by Chandra.  For these
clusters, although interesting X-ray structures are observed in the
cluster cores, no X-ray features related to the motion of the cD
galaxies have been reported \citep{bla01,joh02,kaw03}.

In this paper, we report on Chandra X-ray observations of the remaining
two clusters that have fast moving cDs, A2107 and A2670.  The peculiar
velocities of the cD galaxies are $v_{\mathrm{P}}=433\:\rm km\: s^{-1}$
for A2670, and $270\:\rm km\: s^{-1}$ for A2107 \citep{oeg01}.  For
A2670, Hobbs and Willmore (\yearcite{hob97}) analyzed the existing X-ray
and optical data. While the X-ray morphology on large scales is regular,
the cluster has complicated structures at the center. Hobbs and Willmore
(\yearcite{hob97}) detected four point sources near the center of the
ROSAT HRI image as is shown in their figure~2.  Besides the cD galaxy,
they argued that two of the sources 
(1: \timeform{23h54m09s.5}, \timeform{-10D25'48''}, and 3:
\timeform{23h54m07s.0}, \timeform{-10D25'17''}) were coincident with
cluster galaxies.  The X-ray contours near the cluster center are
elongated towards east and west, which suggests that the cD galaxy and
sources~1 and/or~3 are interacting. \citet{bir94} indicated that the
cluster consists of four subclumps from the spatial and redshift
distributions of the galaxies. The centers of three subclumps (Clumps A,
B, and C) are very close to the cluster center (within a few arcminutes
from the cD galaxy).  This may show that these subclumps are near their
point of closest approach to one another.  \citet{bir94} showed that the
masses of the subclumps are comparable; the subclumps may induce strong
gas motion in the hot intracluster medium (ICM). On the other hand, the
X-ray morphology of A2107 is not reported to be irregular \citep{buo96},
although the detailed X-ray structures in the central region have not
been investigated. Both A2670 and A2107 have relatively weak central
X-ray peaks, which may show that the gas motions in the ICM disrupt the
central gas structures of the clusters. In terms of classical cooling
flows, their peak strengths are equivalent to the mass deposition rates
of $\dot{M}\lesssim 50\: M_\odot \rm yr^{-1}$ \citep{whi97}, although
the classical cooling flow model is no longer valid
\citep{mak01,pet01,kaa01,tam01}.

In the following analysis, we use the cosmological parameters of
$\Omega_0=0.27$, $\lambda=0.73$, and the Hubble constant of $H_0=70\:\rm
km\: s^{-1}\: Mpc^{-1}$ unless otherwise mentioned. This means that
$1\arcsec$ corresponds to 1.45~kpc for A2670 ($z=0.0765$) and 0.813~kpc
for A2107 ($z=0.0411$).

\section{Observations and Data Analysis}

A2670 and A2107 were observed with Chandra on 2004 May 10 for a total of
40~ks and on 2004 September 7 for a total of 36~ks, respectively.  The
data were taken in Very Faint mode using the four ACIS-I chips and the
S2 chip, with the cluster center located on the ACIS-I detector. The
focal plane temperature was $-120$ C, and only events with ASCA grades
of 0, 2, 3, 4, and 6 were analyzed.  We used the data analysis package
CIAO, version 3.1, for the data reductions. Background files were taken
from the blank-sky observations compiled by
M. Markevitch\footnote{http://cxc.harvard.edu/contrib/maxim/bg/} and
included in the CIAO calibration database (CALDB 2.26). We searched for
background flares using data from the S2 chip, since the cluster
emission fills the ACIS-I chips. We used the script {\tt lc\_clean} to
clean the data in the same manner as was done by M. Markevitch for the
blank-sky fields.  After flare removal, the useful exposures were 26~ks
for A2670 and 30~ks for A2107.  The background files were normalized by
requiring the same event count rate as the data in a hard band (PHA
channels between 2500 and 3000).  We did not find any excess soft X-ray
background in the direction of either cluster (i.e., \authorcite{vik05}
\yearcite{vik05}).

\section{X-ray Images}

\begin{figure}
\FigureFile(80mm,80mm){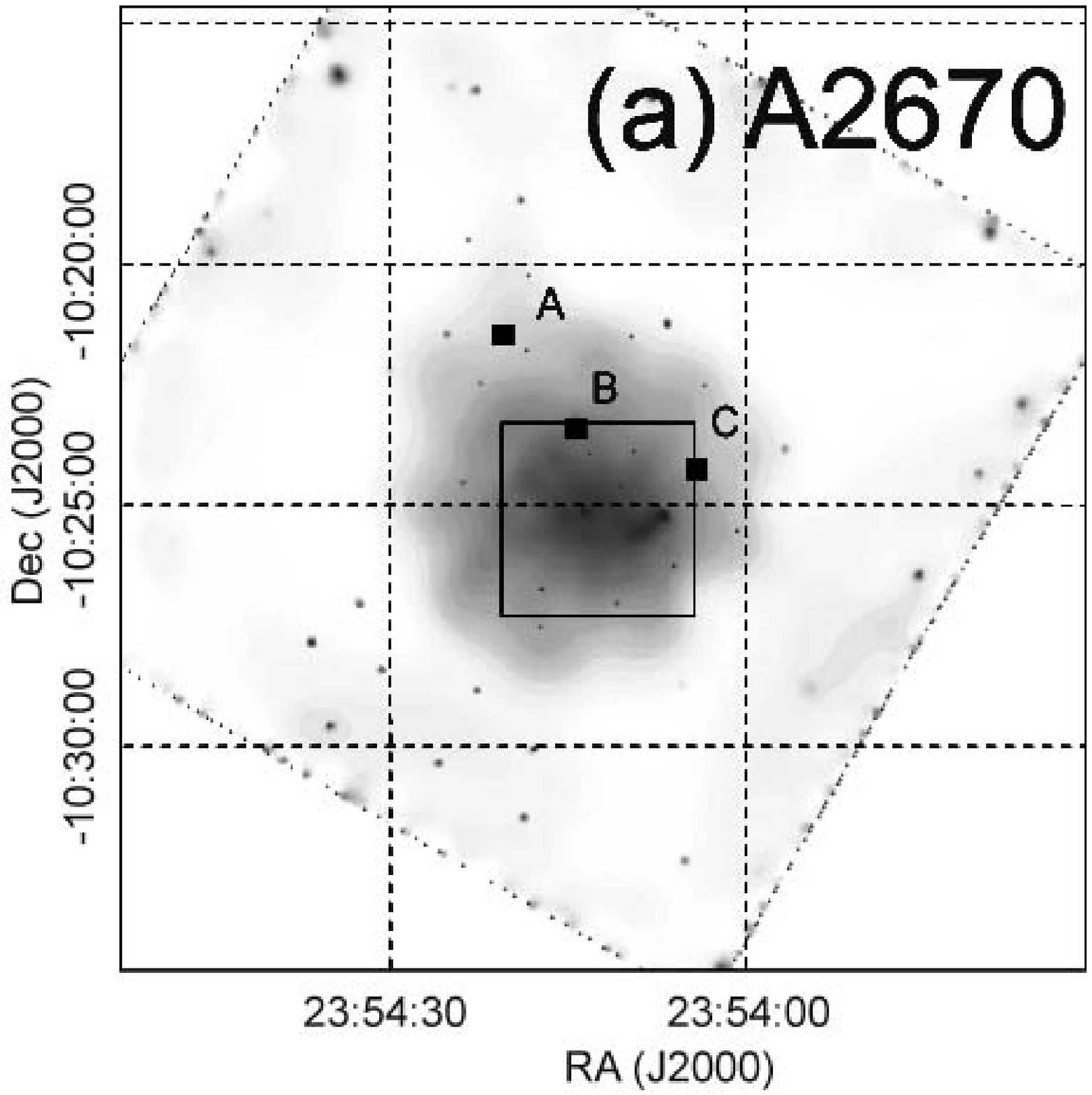} \FigureFile(80mm,80mm){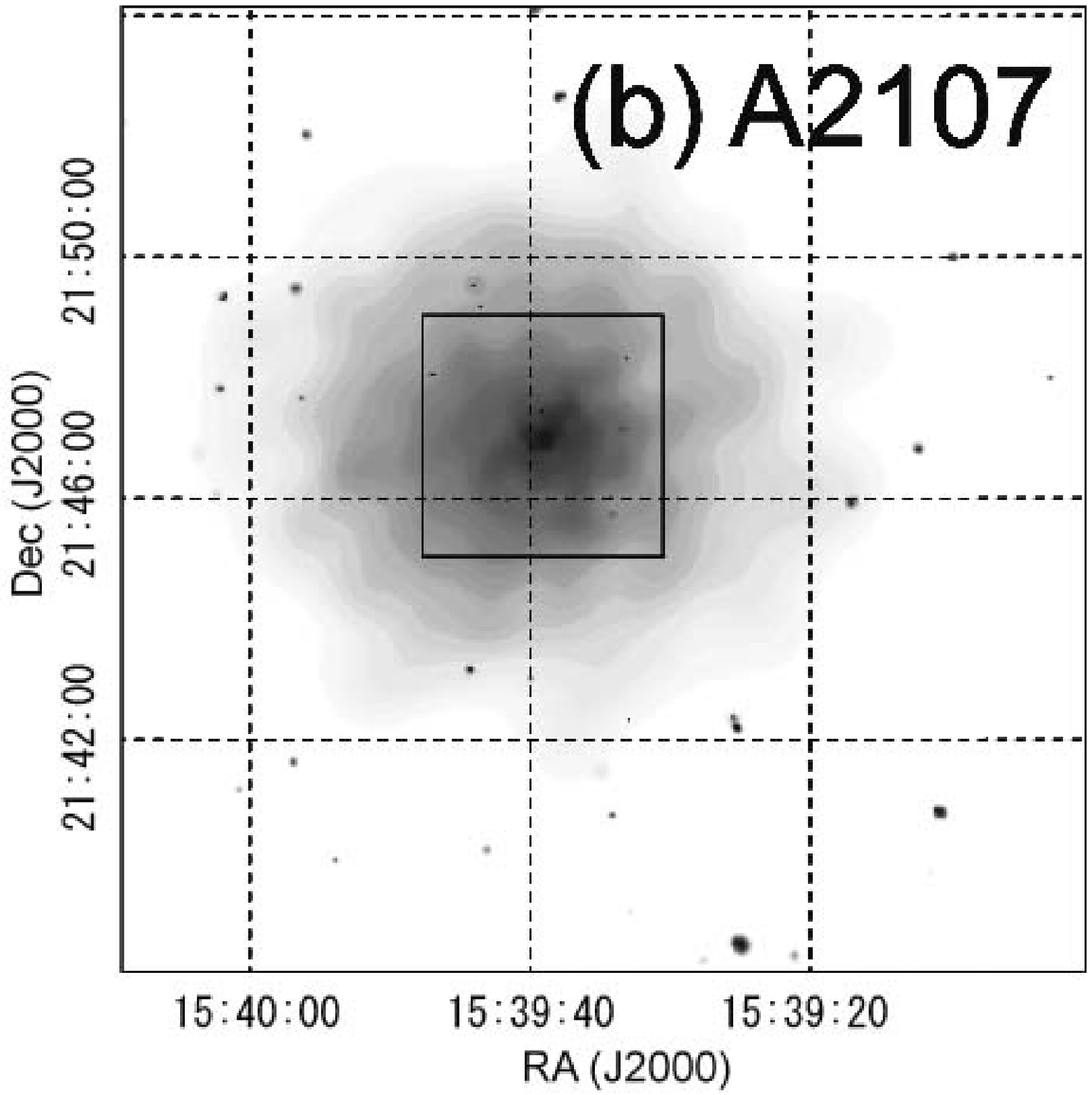}
\caption{Adaptively smoothed, combined ACIS-I images of (a) A2670 and
(b) A2107 in the energy range of 0.3--10 keV, corrected for background,
exposure, and vignetting. The regions shown in figure~\ref{fig:cen} are
indicated as solid-line squares. The filled squares in
figure~\ref{fig:all}a are the positions of the subclumps identified by
\citet{bir94}. Clump~D found by \citet{bir94} is outside this
figure. \label{fig:all}}
\end{figure}

\begin{figure}
\FigureFile(80mm,80mm){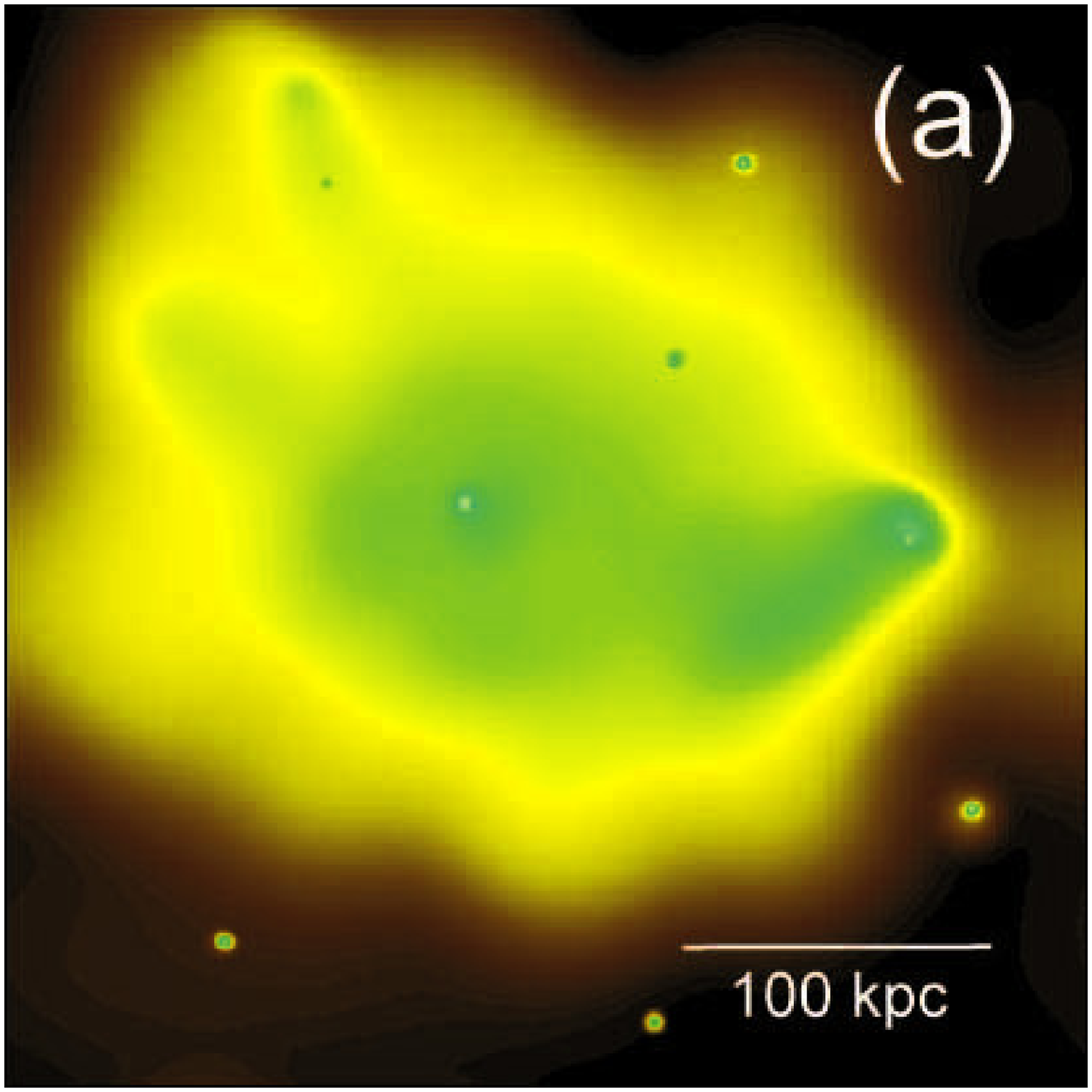}
\FigureFile(80mm,80mm){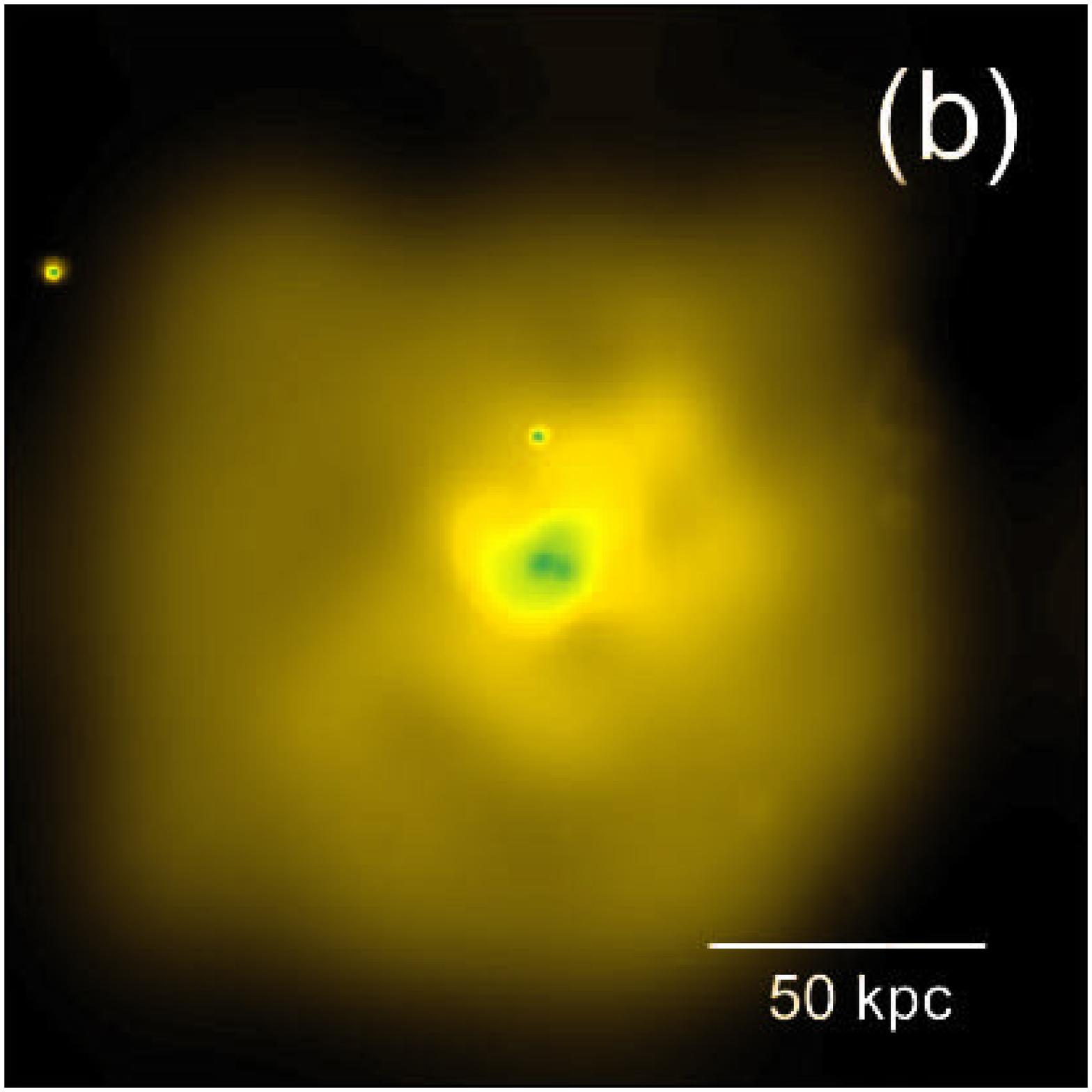} \caption{Adaptively smoothed
Chandra images of the central $4'\times 4'$ regions of (a) A2670
and (b) A2107, corrected for background, exposure, and
vignetting. \label{fig:cen}}
\end{figure}

\begin{figure}
\FigureFile(80mm,80mm){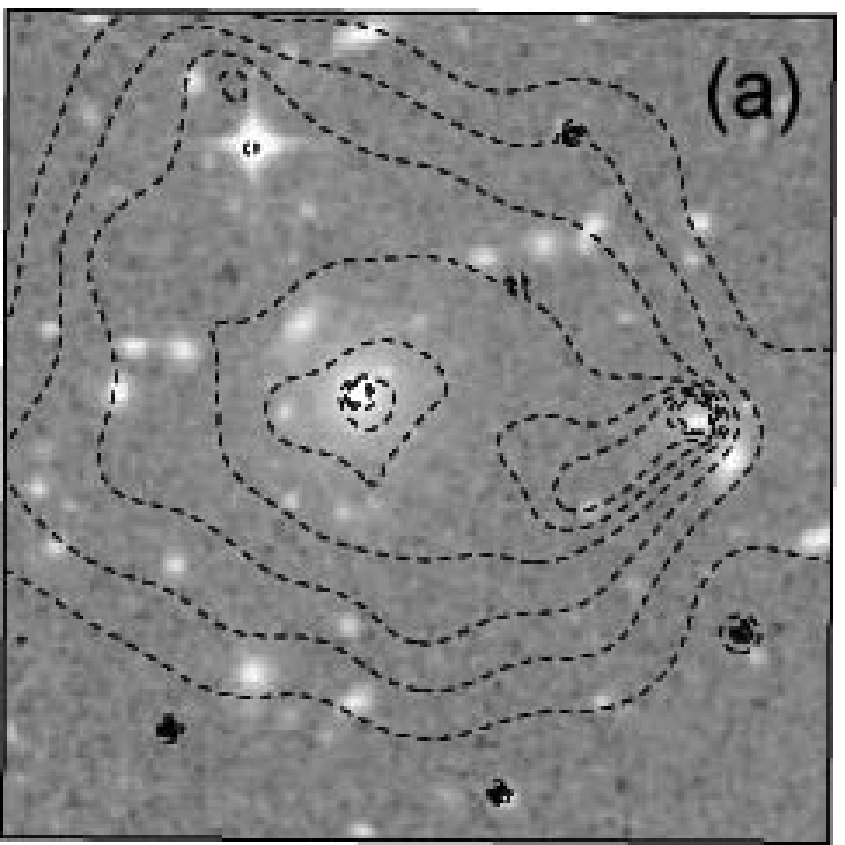}
\FigureFile(80mm,80mm){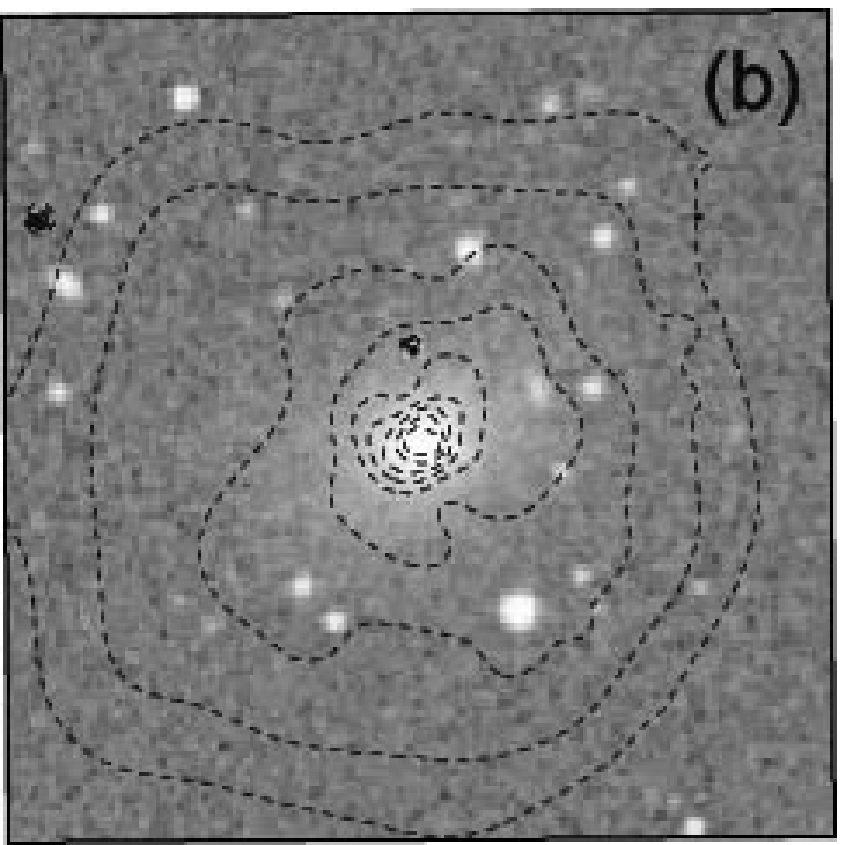} \caption{X-ray brightness
 contours (0.3--10~keV band, logarithmically spaced by a factor of
 $\sqrt{2}$), overlaid on the DSS optical images of the central
 $4'\times 4'$ regions of (a) A2670 and (b) A2107. Contours at the very
 centers of the cDs and comet galaxy are omitted to see the optical
 images. \label{fig:con}}
\end{figure}

Figure~\ref{fig:all} shows the combined ACIS-I images of A2670 and A2107
in the 0.3--10~keV energy band. In figure~\ref{fig:all}a, the positions
of the subclumps in A2670 identified by \citet{bir94} through galaxy
motions are shown. The centers of the clumps are determined by galaxy
positions and the errors may be very large. Figure~\ref{fig:cen} shows
the X-ray images of the central regions of the two clusters. The images
were smoothed using the CIAO routine `csmooth'. The images have a
minimum signal-to-noise ratio of 3 per smoothing beam and were corrected
for exposure, vignetting, and background.  In figure~\ref{fig:con},
contours from the X-ray images are superposed on the Digital Sky Survey
(DSS) images of the two clusters.

In A2670 (figure~\ref{fig:cen}a), the bright source at the center of the
X-ray image corresponds to the cD galaxy near the center of the cluster
(figure~\ref{fig:con}a).  The spatial size of the brightest X-ray region
of the cD galaxy is much smaller than those of cD galaxies in most
cooling core clusters.  The radius is only $\sim 4$~kpc and is smaller
than the optical size of the cD galaxy (figure~\ref{fig:con}a).  The X-ray
source associated with the cD was also seen in the ROSAT HRI image
\citep{hob97}.

In figure~\ref{fig:cen}a, a comet-like X-ray feature is seen to the west
of the cD galaxy in A2670.  The tail of the comet extends from the
elliptical galaxy 2MASX~J23540700--1025169, which is located at the
bright source at the head of the comet (figure~\ref{fig:con}a).  This
X-ray source corresponds with source~3 in figure~2 in Hobbs and Willmore
(\yearcite{hob97}). We will refer to this X-ray comet and the
associated galaxy as the `comet galaxy' from now on.  Although the
leading edge of the comet corresponds to a steep gradient in the X-ray
surface brightness, which we will argue is a cold front
(section~\ref{sec:com_kin} below), there is no clear indication of a bow
shock ahead of the galaxy.  We did not find point sources at the
locations of sources 1 and 2 
(\timeform{23h54m08s.2}, \timeform{-10D25'37''}) found by Hobbs and
Willmore (\yearcite{hob97}) in the ROSAT HRI image.  Instead, it appears
likely that these ``sources'' are just the comet tail in the Chandra
image.  The Chandra image of A2670 also shows an extension to the
northeast of the cD, which might be associated with galaxy groups A and
B in \citet{bir94}.

Compared with A2670, the overall structure of A2107 is more regular
(figure~\ref{fig:all}b).  However, the ICM in the central region
($\lesssim 15\arcsec$ from the center) has an irregular structure
(figure~\ref{fig:cen}b).  This corresponds to the region around the
central cD galaxy (figure~\ref{fig:con}b). One interesting feature is that
the X-ray emission at the center of the cD is elongated roughly
east-west (green color in figure~\ref{fig:cen}b). The optical center of
the cD galaxy is at the center of the elongated structure.

\begin{figure}
\FigureFile(80mm,80mm){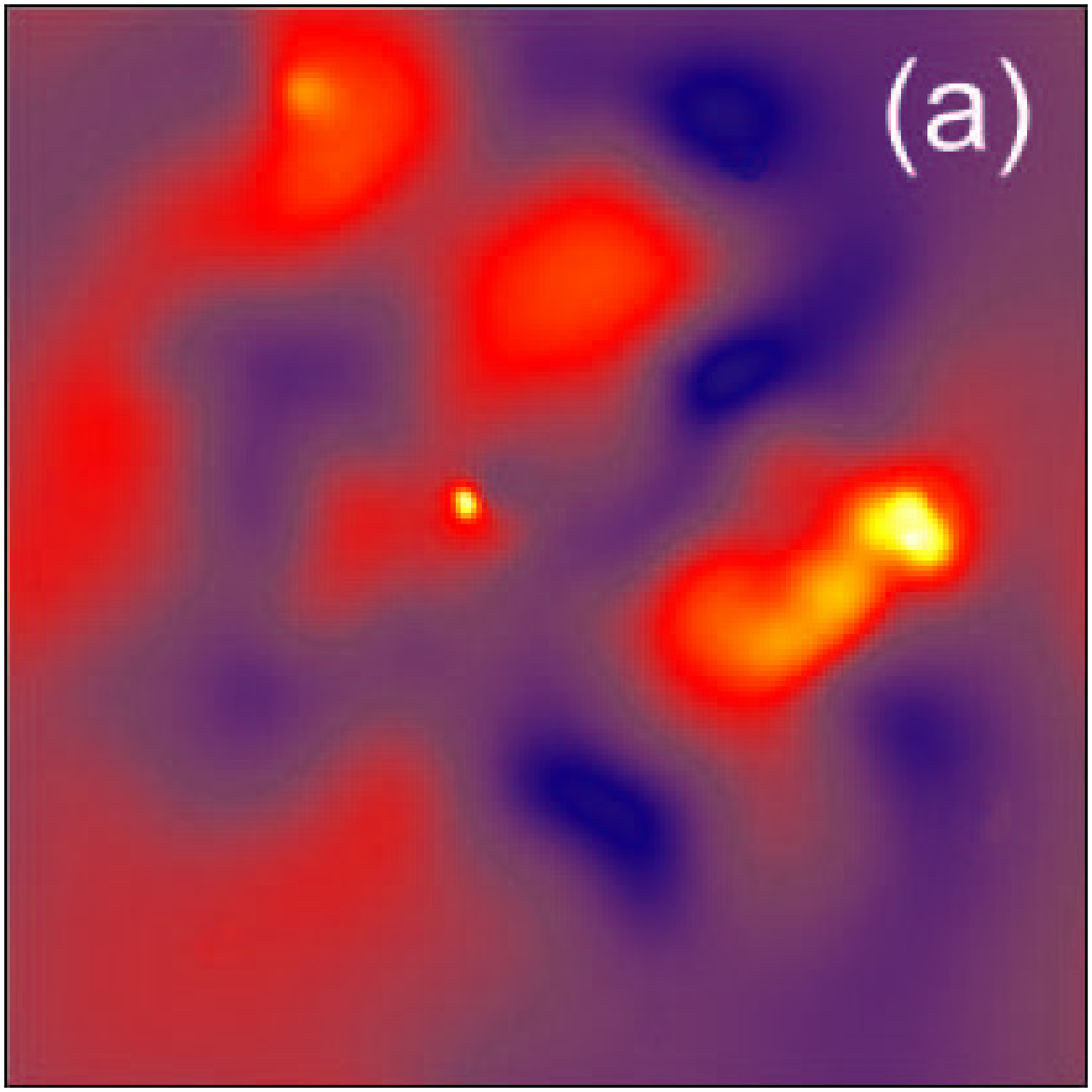}
\FigureFile(80mm,80mm){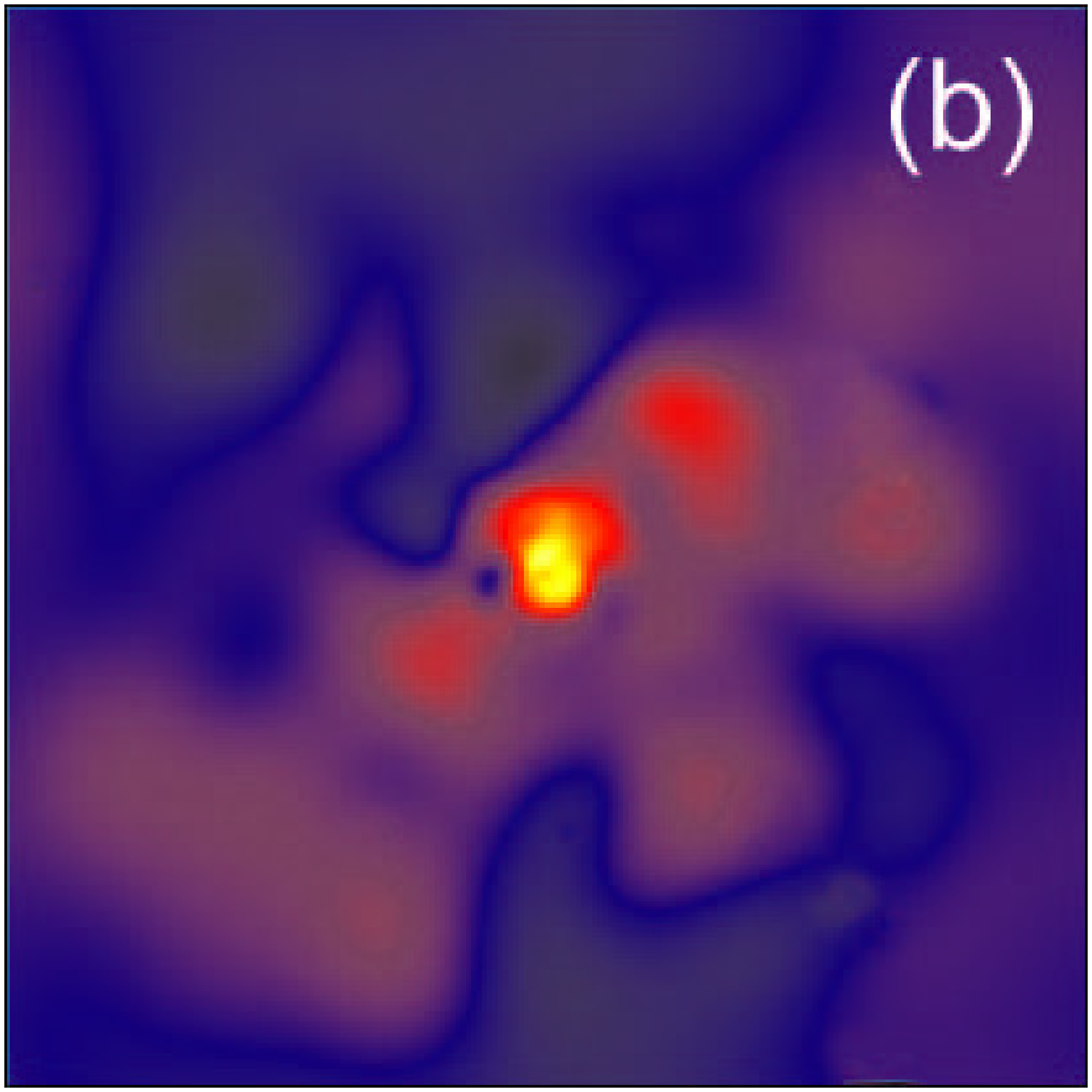} \caption{Color maps of
hardness ratio of the central $4'\times 4'$ regions of (a) A2670 and (b)
A2107, based on the ratio of the count rates in the energy bands 2.0--10
and 0.3--2.0~keV.  The hardness ratios are corrected for background,
exposure, and vignetting and are adaptively smoothed.  Soft emission
appears yellow and hard emission blue. 
Before taking the ratio, the hard and soft images were respectively
smoothed using the same scale map (0.3--10~keV). Note that the color
scales are different between (a) and (b). Deep blue is a hardness ratio
of 0.47 and 0.55, and bright yellow is 0.10 and 0.22 for A2670 and
A2107, respectively. Point sources are removed. \label{fig:HR}}
\end{figure}

Figure~\ref{fig:HR} shows hardness ratio maps of A2670 and A2107, which
are made in order to choose regions for spectral analysis
(section~\ref{sec:spect}). We define the hardness ratio as the count
rate in the 2--10 keV band divided by the count rate in the 0.3--2~keV
band.  The blue color indicates hard X-ray emission, while the yellow
color is the softest emission.  The regions shown are the same as in
figures~\ref{fig:cen} and \ref{fig:con}.

In A2670, the X-ray excess corresponding to the cD galaxy is soft.  The
head of the comet associated with the comet galaxy is also distinctively
soft.  Although it is not as soft as the emission from the two galaxies,
the tail of the comet appears to be softer than the surrounding hot ICM.
The softer X-ray emission from these regions is due to cooler gas than
that in the surrounding cluster (section~\ref{sec:spect} below).  

In the central region of A2107, the structures seen in the hardness
ratio map for the central $\sim 15\arcsec$ region are different from
those seen in the X-ray image (figure~\ref{fig:cen}b).  In the X-ray
image, the emission is elongated east-west,
which is also shown by the innermost contours in figure~\ref{fig:con}b.
Presumably, the X-ray image shows where the densest gas is located.  In
the hardness ratio map (figure~\ref{fig:HR}), the softest (and
presumably coolest) gas is elongated north-south. The difference of
structure between figures~\ref{fig:cen} and~\ref{fig:HR} suggests that
the cluster center is not in pressure equilibrium; the denser gas does
not always correspond to the cooler gas, especially in the azimuthal
direction. Similar structures have been found in the centers of a few
other clusters, including AWM7 \citep{fur03} and 2A0335+096
\citep{maz03}.

\begin{figure}
  \begin{center}
\FigureFile(80mm,80mm){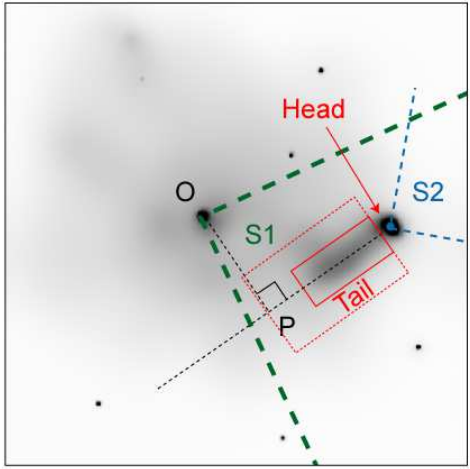} 
  \end{center}
\caption{X-ray image of the center of
A2670, with several regions for further analysis indicated.  The image
is a grey-scale version of figure~\ref{fig:cen}a.  The sectors S1 and S2
are indicated by bold and thin dashed lines, respectively.
\label{fig:region}}
\end{figure}

\begin{figure}
\FigureFile(80mm,80mm){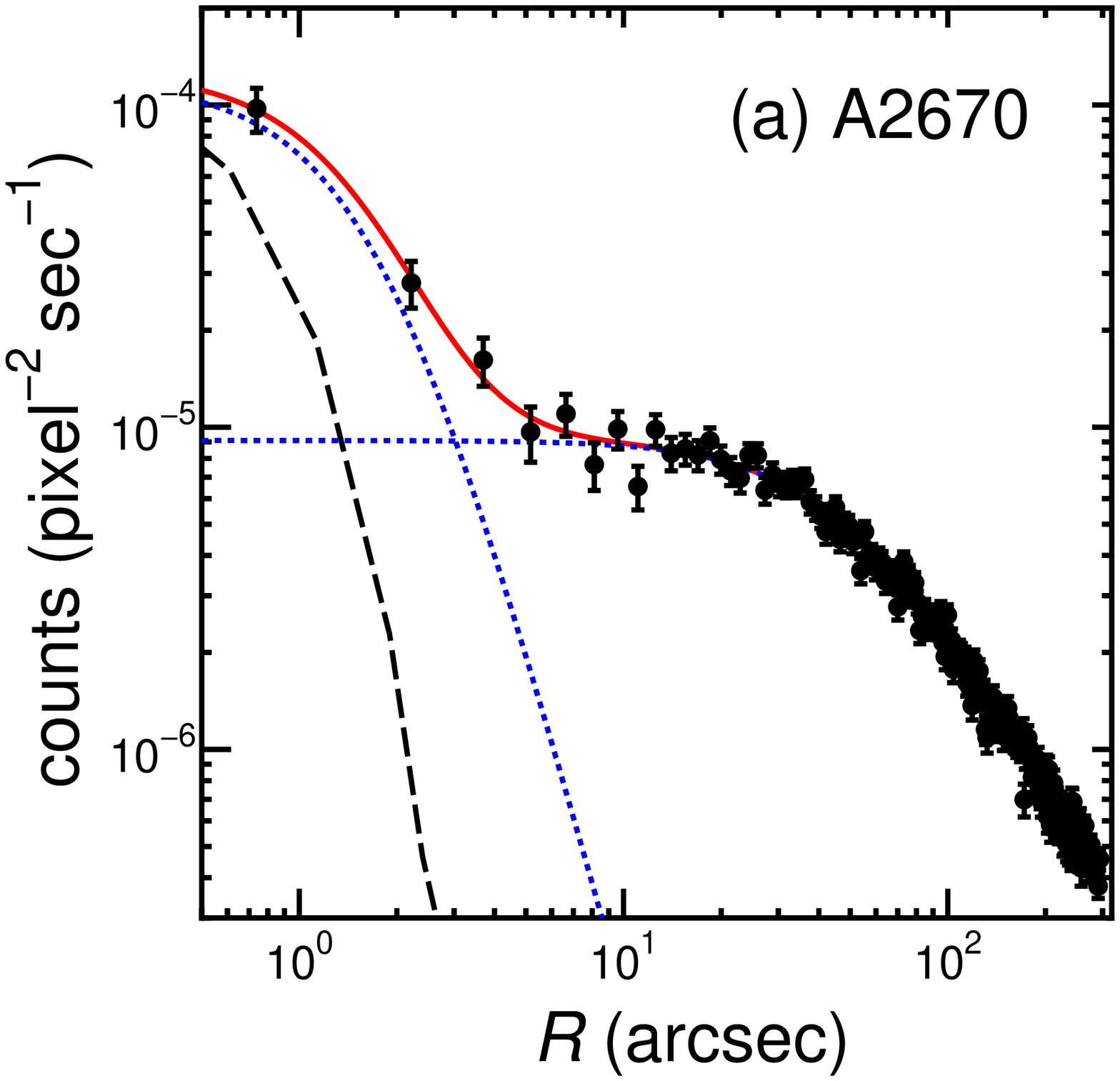} \FigureFile(80mm,80mm){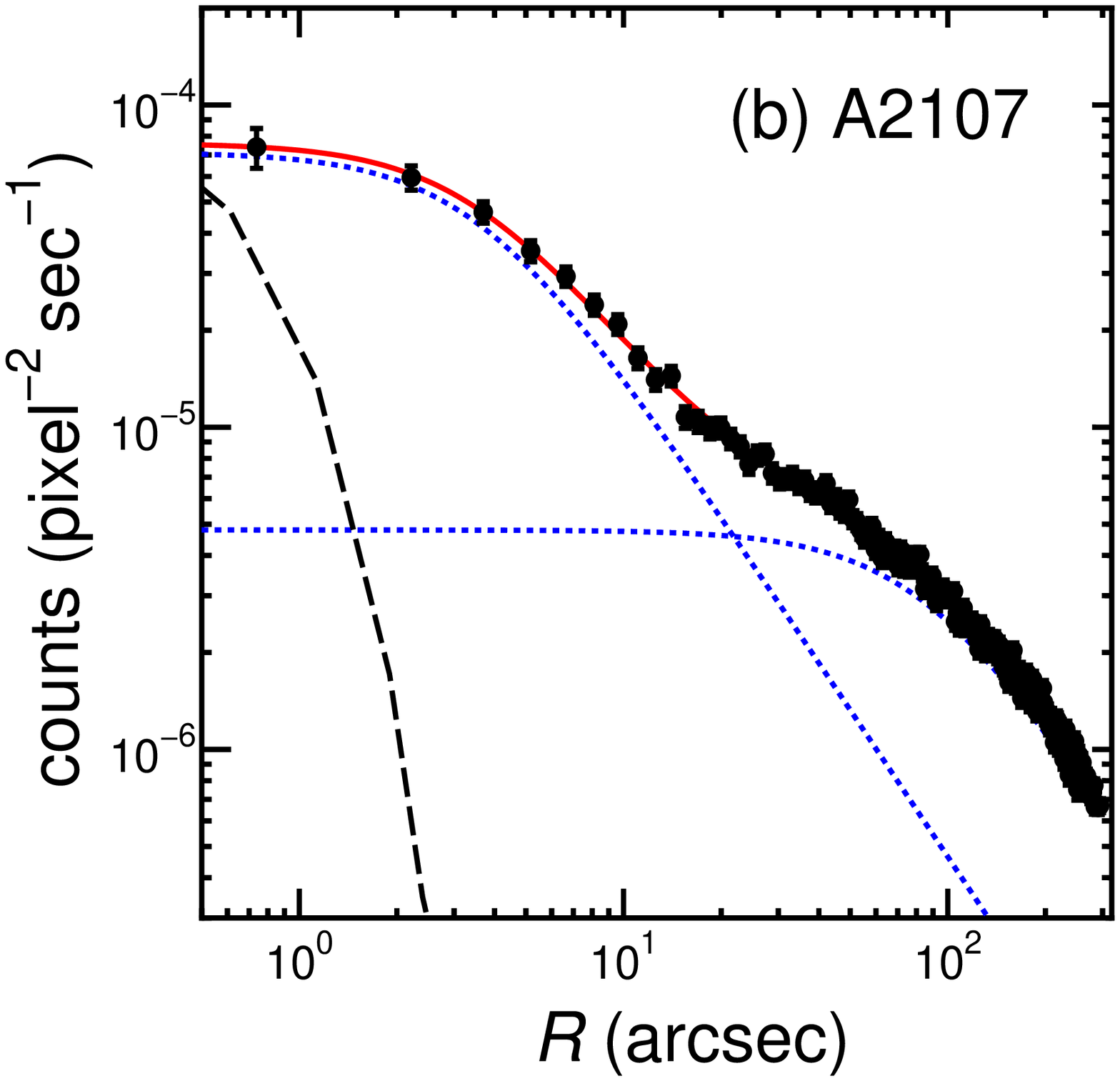}
\caption{Surface brightnesses as a function of radius for (a) A2670
excluding the 90\arcdeg~sector including the comet galaxy (S1 in
figure~\ref{fig:region}) and (b) A2107 for 0.3--10~keV. Error bars are
$1\sigma$ Poisson uncertainties. The results of double $\beta$ model
fits are shown by solid lines. The contribution of each $\beta$ model is
shown by dotted lines. The dashed lines show the expected radial profile
for a point source. \label{fig:surf}}
\end{figure}

The radial X-ray surface brightness profile of each of the clusters was
determined by accumulating counts in circular annuli.  The profiles were
centered on the peaks in the X-ray surface brightness coincident with
the X-ray centers of the cD galaxies.  For A2670, the 90\arcdeg~sector
including the comet galaxy (S1 in figure~\ref{fig:region}) was excluded
from the surface brightness profile.  The counts in annuli were
corrected for background, vignetting, and exposure.  The resulting
surface brightness profiles are shown in figure~\ref{fig:surf}.  We
first tried to fit a single $\beta$ model to the surface brightness
profiles, but the fits were not acceptable.  Figure~\ref{fig:surf}
suggests that there are two components to the surface brightness of each
cluster.  Thus, we fit the radial profiles of the surface brightness
with double $\beta$ model functions
\begin{equation}
\label{eq:beta}
 S_X(R)=S_{0,1}\left(1+\frac{R^2}{r_{cl}^2}\right)^{-3\beta_1+1/2}
+S_{0,2}\left(1+\frac{R^2}{r_{c2}^2}\right)^{-3\beta_2+1/2}\:,
\end{equation}
where $R$ is the projected distance from the cluster center.  This fit
yields values for the core radii $r_{c1}$ and $r_{c2}$, the $\beta$
parameters $\beta_1$ and $\beta_2$, and the normalizations $S_{0,1}$ and
$S_{0,2}$. We adopted the double $\beta$ model functions because they
are easily converted into a density profile (see
equation~[\ref{eq:rhogas_ob1}]). 
The best-fit parameters are given in table~\ref{tab:beta}, and the
best-fit total model profiles and individual components are shown in
figure~\ref{fig:surf}.

\begin{table}
{\scriptsize
\begin{center}
\caption{Parameters for $\beta$ model fits \label{tab:beta}}
 \begin{tabular}{lcccccc}
  \hline\hline
  \multicolumn{1}{c}{Cluster} &
$r_{c1}$ &
$r_{c2}$ &
$\beta_1$ &
$\beta_2$ &
$S_{0,1}$ &
$S_{0,2}$ \\
 &
(kpc) &
(kpc) &
 &
 &
($10^{-5}\rm cnts\: s^{-1} pix^{-2}$) &
($10^{-5}\rm cnts\: s^{-1} pix^{-2}$) 
\\
  \hline
A2670 (Excluding Comet\footnotemark[$*$])&
$2.55_{-1.28}^{+1.28}$&
$66.0_{-3.2}^{+3.2}$  &
$0.79_{-0.29}^{+0.29}$&
$0.44_{-0.01}^{+0.01}$&
$11.8_{-3.3}^{+3.3}$  &
$0.91_{-0.03}^{+0.03}$
\\
A2670 (Comet Direction\footnotemark[$\dagger$])&
11.3\footnotemark[$\ddagger$]&
...&
$0.49_{-0.01}^{+0.01}$&
...&
$38.3_{-8.5}^{+8.5}$&
...
\\
A2170 &
$3.15_{-0.67}^{+0.67}$&
$78.8_{-11.7}^{+11.7}$&
$0.42_{-0.04}^{+0.04}$&
$0.44_{-0.02}^{+0.02}$&
$7.14_{-0.81}^{+0.81}$&
$0.48_{-0.01}^{+0.01}$
\\
  \hline
 \multicolumn{4}{@{}l@{}}{\hbox to 0pt{\parbox{130mm}{\footnotesize
  ~ 
  \par\noindent
  \footnotemark[$*$] Excluding the 90\arcdeg~sector including
the comet galaxy (outside S1 in Fig.~\ref{fig:region}).
  \par\noindent
  \footnotemark[$\dagger$] The 90\arcdeg~sector including 
the comet galaxy (S1 in
Fig.~\ref{fig:region}, 
see section~\ref{sec:comet}).  
Single $\beta$ model fit.
  \par\noindent
  \footnotemark[$\ddagger$] Fixed.
}\hss}}
  \end{tabular}
\end{center}
}
\end{table}

The core radii of the inner components of both clusters are very small
($\sim$3 kpc). This is much smaller than any characteristic scale
expected to be associated with a cluster of galaxies, and thus we will
assume that these components are associated with the central cD
galaxies.  Even with that assumption, these core radii are small.  On
the other hand, the central components are clearly resolved; thus, they
are not point sources associated with central AGNs in the cD galaxies.
The dashed lines in figure~\ref{fig:surf} show the point-spread-function
for a source with a spectrum similar to that of the central emission.  A
careful examination of the raw data images shows no evidence for an
unresolved point source at the center of either cD galaxies.  Thus, it
is likely that these central components are emission from highly
compressed gas.  The outer core radii are similar to the sizes of
cooling cores seen in relaxed clusters.

\section{Spectral Analysis}
\label{sec:spect}

We extracted spectra in the 0.3--10~keV band in PI channels from
selected regions of the Chandra images using the CIAO software
package. Both the response matrix files and the ancillary response
files were calculated using the MKRMF/MKWARF package.  The package
weighted the response files by the X-ray brightness over the
corresponding image region. The spectra were grouped to have a minimum
of 20 counts per bin and fitted with one or two thermal models using
XSPEC, version 11.3.0. As a thermal model, we use APEC, unless otherwise
mentioned. 
We also use WABS for absorption ($N_H$). Errors on fitted spectral
parameters are given at the 90\% confidence level in this section.

\subsection{Global Spectrum}

First, we studied the average spectra extracted from circular regions of
4\farcm92 radius centered on the cD galaxies.  The spectra can be
reproduced by a single thermal component with absorption (1T models in
table~\ref{tab:sp}). For A2670, the temperature $T$ is consistent with
previous ROSAT and ASCA observations ($3.7_{-0.3}^{+0.3}$~keV;
\authorcite{hob97} \yearcite{hob97}).  The metal abundance obtained with
Chandra (table~\ref{tab:sp}) is larger than that obtained with ASCA
($Z=0.20_{-0.09}^{+0.11}\: Z_\odot$; \authorcite{hob97}
\yearcite{hob97}). One possible reason could be that the cluster has a
metal abundance excess at the center like many other clusters
\citep{fuk00}; since the field of view of Chandra is smaller than ASCA,
the metal abundance measured with Chandra could be larger than that
measured with ASCA. The absorption we obtain with Chandra
(table~\ref{tab:sp}) is larger than the Galactic value ($N_{\rm
H}=2.92\times 10^{20}\rm\: cm^{-2}$; \authorcite{sta92}
\yearcite{sta92}) and that obtained with ROSAT ($N_{\rm H}=3.6\times
10^{20}\rm\: cm^{-2}$; \authorcite{hob97} \yearcite{hob97}), although it
is consistent with the ASCA result ($N_{\rm H}=6_{-5}^{+6}\times
10^{20}\rm\: cm^{-2}$; \authorcite{hob97} \yearcite{hob97}). 
The excess absorption exists even if we use ACISABS software. If the
ROSAT result is correct, one possible reason of the discrepancy might be
an imperfect correction of the quantum efficiency degradation of the
ACIS detector on Chandra at low energies.  Because of this possible
source of uncertainty, we do not discuss $N_{\rm H}$ further.  For
A2107, the temperature obtained with Chandra is consistent with that
obtained with Einstein MPC (4.1~keV; \authorcite{dav93}
\yearcite{dav93}). The best-fit absorption is larger than the Galactic
value ($N_{\rm H}=5.02\times 10^{20}\rm\: cm^{-2}$; \authorcite{sta92}
\yearcite{sta92}).

\begin{table}
{\scriptsize
\begin{center}
\caption{X-ray Spectra of A2670 and A2107 \label{tab:sp}}
 \begin{tabular}{lccccccc}
  \hline\hline
 &
Radius &
$T$ &
$Z$ &
$T_{\rm out}$ &
$N_{\rm H}$ &
$\dot{M}$ &
$\chi^2$/dof \\
Region (Model) &
 &
(keV) &
($Z_{\odot}$) &
(keV) &
($10^{20}\rm\; cm^{-2}$) &
($M_\odot\:\rm yr^{-1}$) & 
\\
  \hline
A2670, Global (1T) &
	4\farcm92 &
	$3.8_{-0.2}^{+0.2}$ &
	$0.47_{-0.08}^{+0.09}$ &
	... &
	$7.1_{-1.0}^{+1.0}$ &
	... &
	0.979 (345.3/401) \\
A2670, Global (CF) &
	4\farcm92 &
	$3.9_{-0.2}^{+0.2}$ &
	$0.42_{-0.07}^{+0.08}$ &
	... &
	$6.9_{-0.9}^{+1.2}$ &
	$0_{-0}^{+11.7}$ &
	0.980 (343.8/400) \\
A2670, Comet Head (2T\footnotemark[$*$] ) &
	8\farcs9 &
	$1.4_{-0.1}^{+0.3}$ &
	0.47\footnotemark[$\dagger$]  &
	4.7\footnotemark[$\ddagger$] &
	7.1\footnotemark[$\dagger$]  &
	... &
	0.570 (11.4/20) \\
A2670, Comet Head (2T\footnotemark[$*$] ) &
	8\farcs9 &
	$1.6_{-0.3}^{+0.2}$ &
	1.0\footnotemark[$\dagger$]  &
	4.7\footnotemark[$\ddagger$] &
	7.1\footnotemark[$\dagger$]  &
	... &
	0.512 (10.2/20) \\
A2670, Comet Tail (2T\footnotemark[$*$] ) &
	$49\arcsec\times 22\arcsec$ &
	$1.5_{-0.3}^{+0.6}$ &
	0.47\footnotemark[$\dagger$]  &
	3.6\footnotemark[$\ddagger$] &
	7.1\footnotemark[$\dagger$]  &
	... &
	0.631 (29.0/46) \\
A2670, Comet Tail (2T\footnotemark[$*$] ) &
	$49\arcsec\times 22\arcsec$ &
	$1.7_{-0.4}^{+0.5}$ &
	1.0\footnotemark[$\dagger$]  &
	3.6\footnotemark[$\ddagger$] &
	7.1\footnotemark[$\dagger$]  &
	... &
	0.602 (27.7/46) \\
A2107, Global (1T) &
	4\farcm92 &
	$4.0_{-0.1}^{+0.1}$ &
	$0.44_{-0.06}^{+0.07}$ &
	... &
	$7.9_{-0.7}^{+0.7}$ &
	... &
	0.896 (467.7/522) \\
A2107, Global (CF) &
	4\farcm92 &
	$4.1_{-0.1}^{+0.2}$ &
	$0.39_{-0.06}^{+0.06}$ &
	... &
	$7.8_{-0.8}^{+0.9}$ &
	$1.1_{-1.1}^{+3.8}$ &
	0.891 (464.0/521) \\
  \hline
 \multicolumn{4}{@{}l@{}}{\hbox to 0pt{\parbox{85mm}{\footnotesize
  ~
  \par\noindent
  \footnotemark[$*$] Deprojection.
  \par\noindent
  \footnotemark[$\dagger$] Assumed.
  \par\noindent
  \footnotemark[$\ddagger$] Fixed.
}\hss}}
  \end{tabular}
\end{center}
}
\end{table}

We also fitted the global spectra with the MKCFLOW model based on the
MEKAL thermal model in order to estimate the emission from low
temperature gas.  Although this model may not give actual cooling rates
if the standard cooling flow is wrong, it should give cooling rates at
least qualitatively. Thus, the results would be useful to be compared
with those of previous studies. We added a MEKAL model representing the
emission outside the cooling core and a variable WABS absorption
model
($N_H$ in table~\ref{tab:sp}). We fix the metal abundance and initial
gas temperature in the MKCFLOW component to the values of abundance and
temperature of the MEKAL component, respectively. The results are shown
in table~\ref{tab:sp} as CF (for `cooling flow') models. The mass
deposition rates, $\dot{M}$, are very small and consistent with zero. We
note that the results are not sensitive to the metal abundance of the
MKCFLOW component; even if we assume that $Z=1\: Z_\odot$ for the
MKCFLOW component and if $Z$ is not fixed for the MEKAL component, the
results of spectral fits are almost the same.

\subsection{Individual Regions}

Next, we studied spectra of individual regions. Unfortunately, total
photon counts for both clusters are relatively small. Therefore, we
only investigated the spectra of several interesting regions and fixed
the metal abundance and the absorption at the values for the innermost
4\farcm92 radius (1T models in table~\ref{tab:sp}) unless otherwise
mentioned.

\subsubsection{Temperature and Density Profiles}

The hardness ratio maps show that the cluster centers (cD galaxies) are
cooler than the surrounding regions (figure~\ref{fig:HR}). Therefore, the
effect of projection of the hotter outer gas onto the cooler central
regions must be considered for the spectral analysis.

Based on the hardness ratio maps, we extracted the spectra from a few
annular regions.  In A2670, these annuli had projected radii of
$3\arcsec<R<260\arcsec$ (region~1) and $0\arcsec<R<3\arcsec$ (region~2).
The projected radii were $15\arcsec<R<260\arcsec$ (region~1),
$7\arcsec<R<15\arcsec$ (region 2), and $0\arcsec<R<7\arcsec$ (region~3)
for A2107.  (Further division of the regions gives spectra with too few
counts to give meaningful constraints on the temperatures.)  We assumed
that the clusters are spherically symmetric, although the
90\arcdeg~sector including the comet galaxy (S1 in
figure~\ref{fig:region}) is excluded for A2670. The spherical shells
corresponding to the above annuli are numbered in a corresponding way.

Usual deprojection analysis assumes that both density and temperature
within each spherical shell are constant \citep{bla01,fuj02}. The
hardness ratio map indicates that the assumption is not bad for
temperature (figure~\ref{fig:HR}). On the other hand, it is not
appropriate for density, because the number of the shells is too small
and thus the widths of the individual shells are too large to assume
that. Therefore, we modified the usual deprojection analysis in the
following way. In this analysis, we assumed that the temperature within
each spherical shell is constant, but the density is not. 

We assume that the X-ray surface brightness profile is given by the
double $\beta$ model fits in table~\ref{tab:beta}. If the temperature
is constant throughout the cluster, the double $\beta$ model can be
deprojected to give the density profile
\begin{equation}
\label{eq:rhogas_ob1}
n_{{\rm gas}, i}(r)^2=n_{{\rm gas}, 1, i}^2
\left(1+\frac{r^2}{r_{c1}^2}\right)^{-3\beta_1}
+ n_{{\rm gas}, 2, i}^2
\left(1+\frac{r^2}{r_{c2}^2}\right)^{-3\beta_2} \:,
\end{equation}
where $r$ is the radius from the cluster center.  Here, since the
temperature is assumed to be different in each shell, we assume
the same form applies but with a different normalization for each shell
$i$. 
That is, the parameters $r_{c1}$, $r_{c2}$, $\beta_1$, and $\beta_2$,
are not dependent on the shell $i$, but $n_{{\rm gas}, 1, i}$ and
$n_{{\rm gas}, 2, i}$ are. The normalization is chosen to be consistent
with the observed surface brightness (table~\ref{tab:beta}) and the
emissivity resulting from the temperature in that shell.
Thus, a temperature and a normalization are assigned to each shell in
our procedure. (On the other hand, a temperature and an uniform density
are assigned to each shell in the conventional deprojection method.) In
other words, instead of the uniform density, we used the normalization
of the density profile, fixing the functional form (the double $\beta$
model). The individual normalizations $n_{{\rm gas}, 1, i}$ and $n_{{\rm
gas}, 2, i}$ can be determined from the overall normalization of each
shell, and the relation
\begin{equation}
\label{eq:rho1}
n_{{\rm gas}, 2, i} =
\left(\frac{S_{0,2} r_{c1}}{S_{0,1} r_{c2}}\right)^{1/2}
n_{{\rm gas}, 1, i} \:.
\end{equation}
%
Thus, the normalization is given by one parameter ($n_{{\rm gas}, 1,
i}$).  Note that although we use the results of double $\beta$ model
fits, we do not assume two-phase structure to the gas (two temperatures
at a given radius).  We adopted equation~(\ref{eq:rhogas_ob1}) just as a
functional form for the X-ray surface brightness profile.  

To include projection effects on the spectral fits, we first fitted the
spectrum of region~1 with one thermal model. Using the result of this
spectral fit, we obtain the normalization of the density profile for
shell~1 ($n_{{\rm gas}, 1, 1}$).  Then, we determined the projected
spectrum of shell~1 onto region~2.  We fit the spectrum of region~2 with
two thermal models; one is the projection of shell~1, which has been
derived using equation~(\ref{eq:rhogas_ob1}) and is fixed, and the other
is the spectrum of shell~2. From the result of the spectral fit, we can
obtain the temperature and the density profile in shell~2 ($i=2$ in
equation~[\ref{eq:rhogas_ob1}]).  For A2107, we repeat this procedure
for shell~3.

\begin{figure}
\FigureFile(80mm,80mm){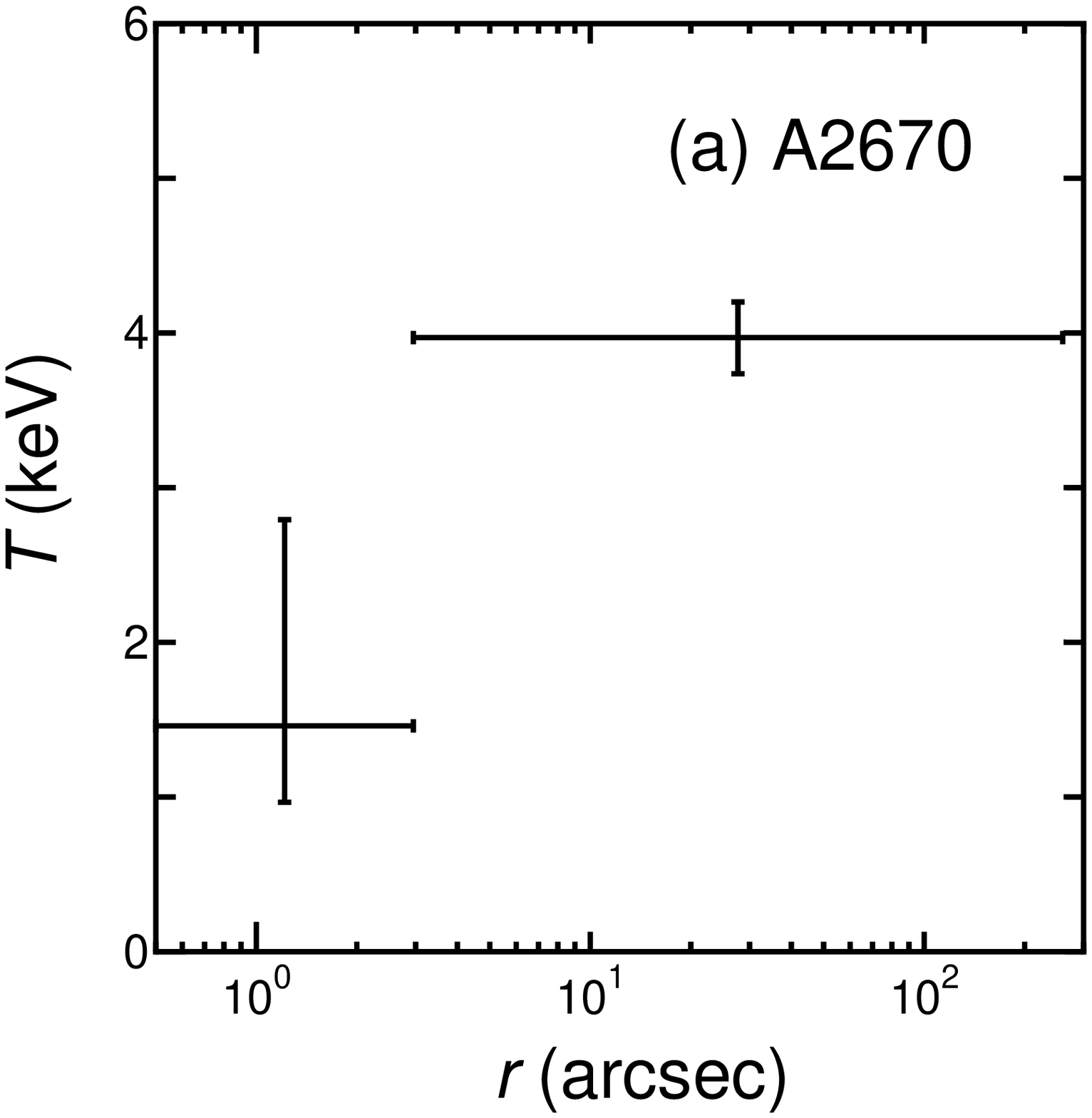}
\FigureFile(80mm,80mm){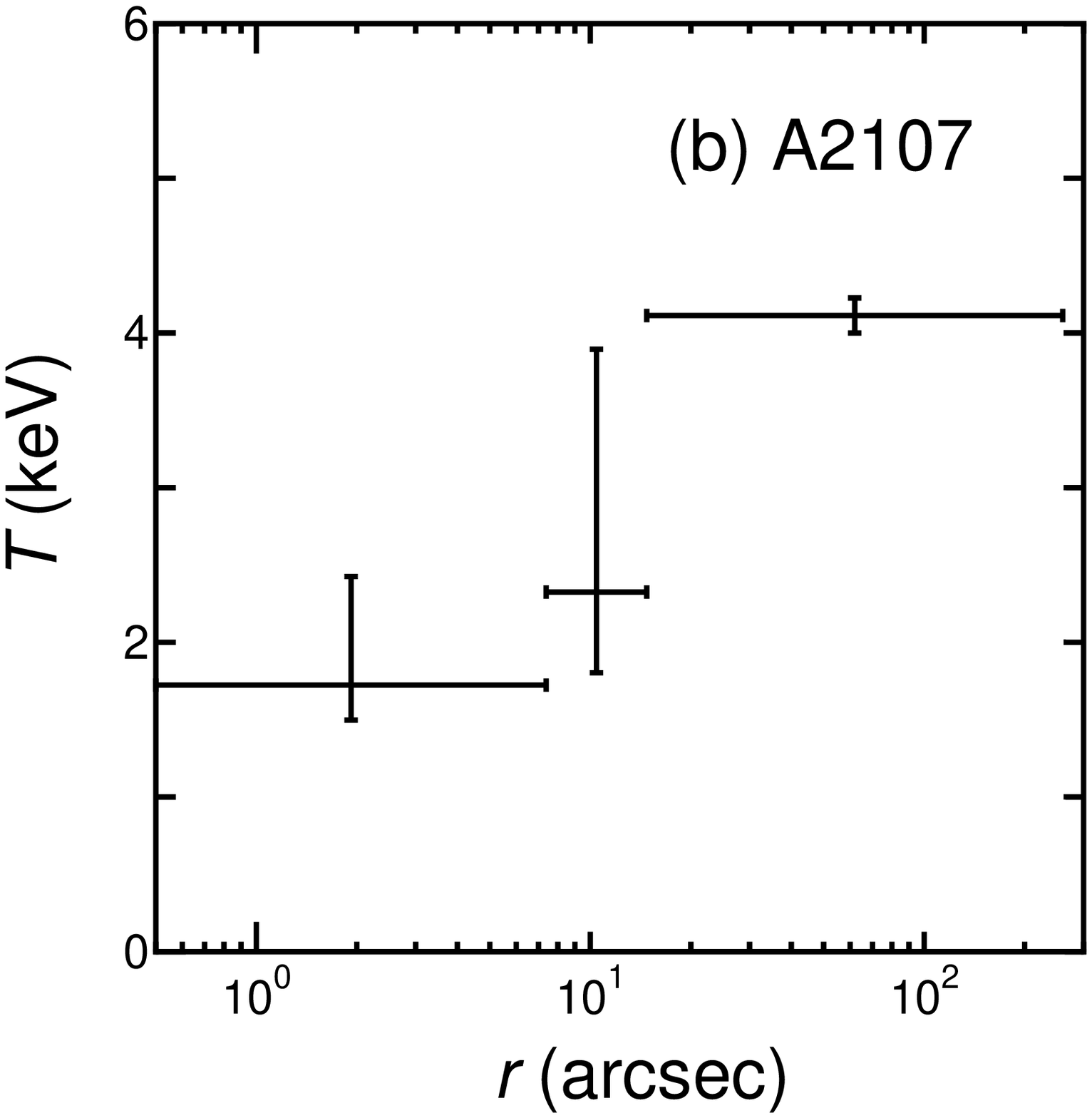} \caption{Temperature as a
function of radius for (a) A2670 excluding the 90\arcdeg~sector
including the comet galaxy (S1 in figure~\ref{fig:region}) and (b) A2107.
\label{fig:T}}
\end{figure}

\begin{figure}
\FigureFile(80mm,80mm){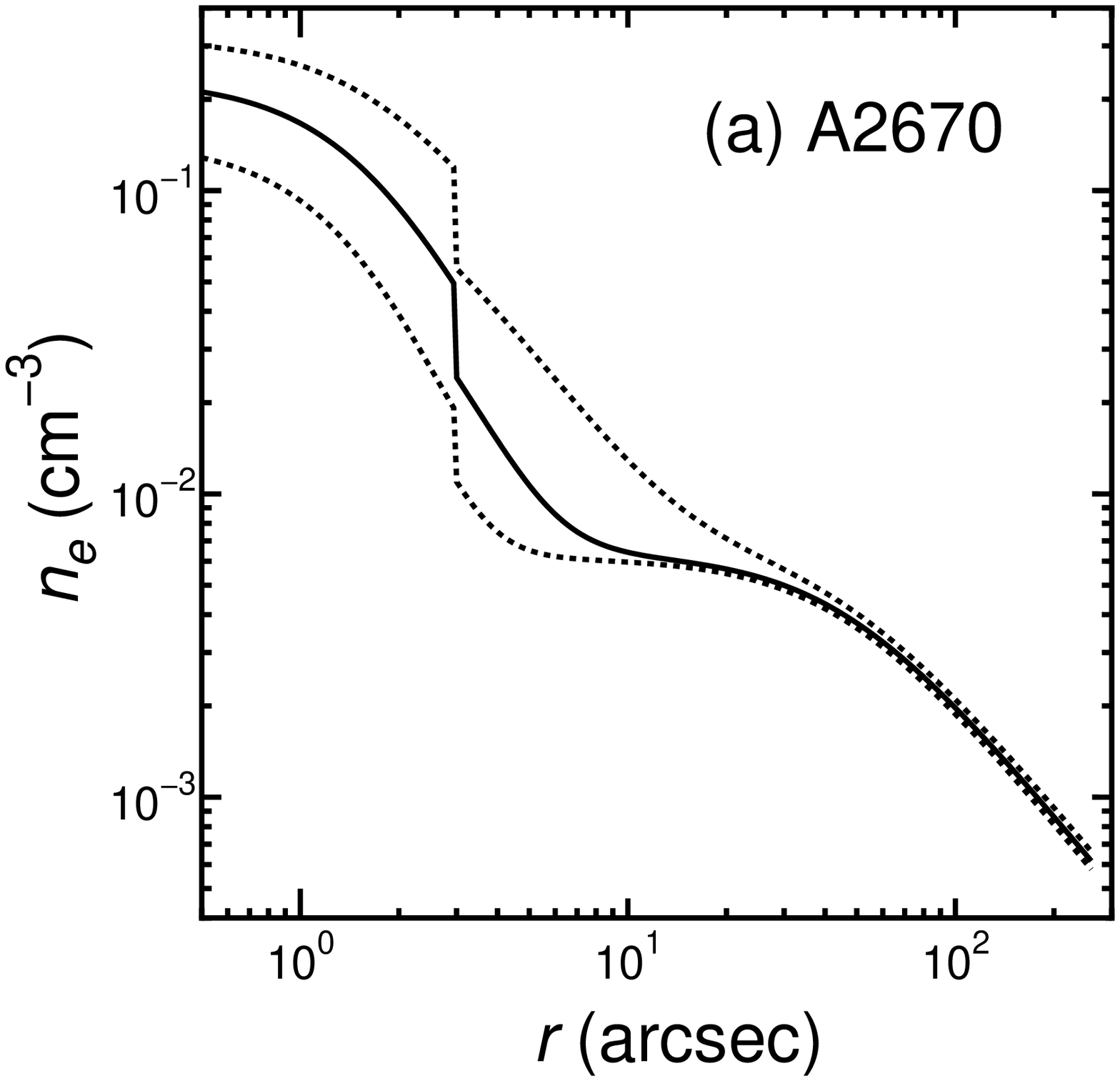} \FigureFile(80mm,80mm){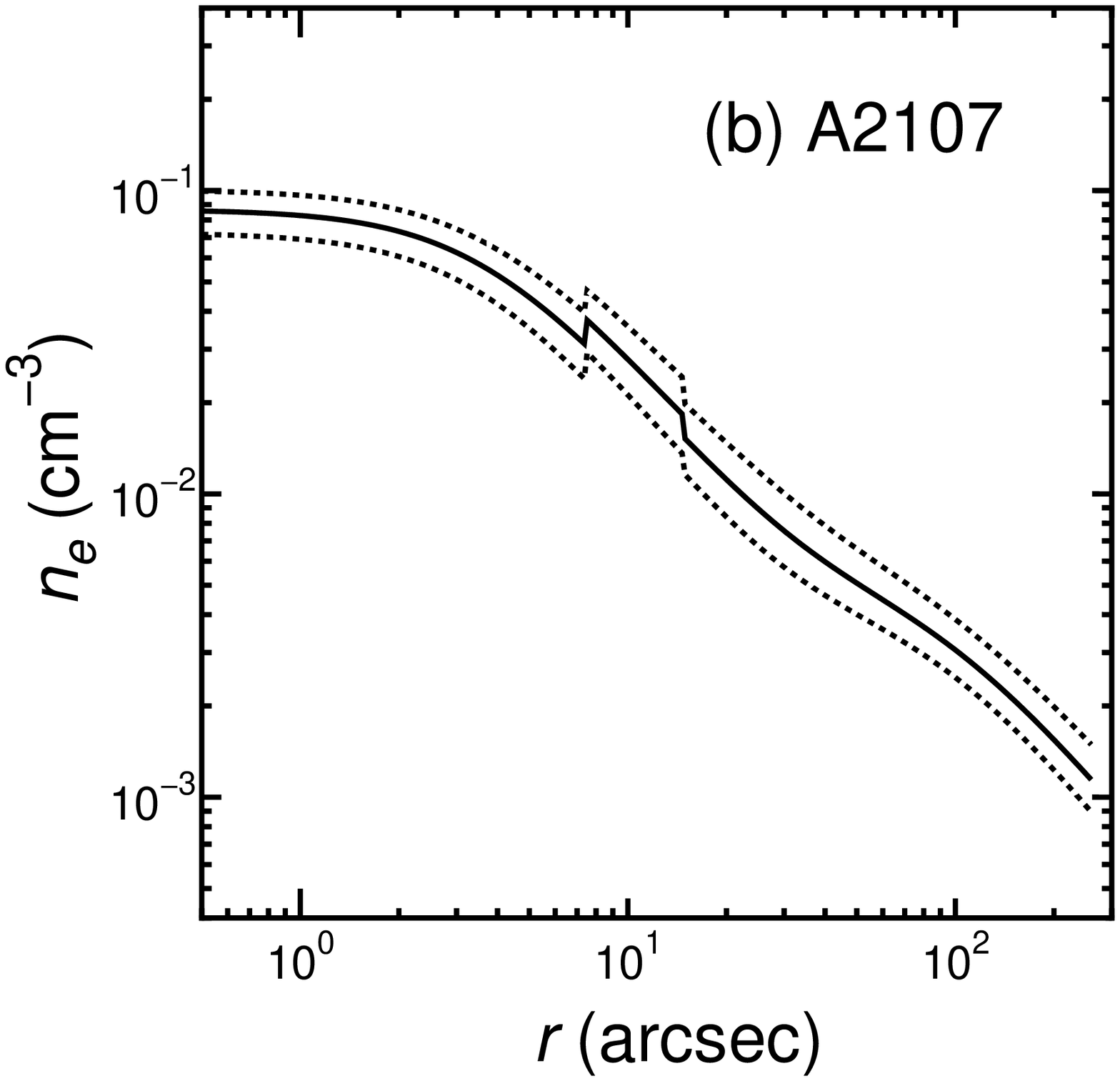}
\caption{Electron density as a function of radius for (a) A2670
excluding the 90\arcdeg~sector including the comet galaxy (S1 in
figure~\ref{fig:region}) and (b) A2107. Errors are shown by dotted
lines,
and they are obtained by changing the parameters for $\beta$ model fits
from their best-fit values (table~\ref{tab:beta}); the errors in the
normalization of a thermal model (APEC) in spectral fits are also
considered. The discontinuities in the figures are artifacts due to the
finite regions used to determine the temperature.  \label{fig:ne}}
\end{figure}

Figures~\ref{fig:T} and~\ref{fig:ne} show the temperature and density
profiles for A2670 and A2107 including projection effects. The cluster
centers (cD galaxies) are cold.  The discontinuities in the density
profiles correspond to the boundaries of the regions for which the
spectral analysis is done, and are artifacts.  The error of density in
the inner most region of A2670 ($<3''$) is relatively large
(figure~\ref{fig:ne}a) because of the uncertainties of the $\beta$ model
fit. If we assume that the density in the region is constant as is
assumed in the usual deprojection analysis, the error should be smaller.

\subsubsection{The Comet Galaxy}
\label{sec:comet}

The spectrum of the comet galaxy in A2670 is also affected by the
projected spectrum of the outer regions of the cluster. First, we fit
the surface brightness of the 90\arcdeg~sector including the comet
galaxy (S1 in figure~\ref{fig:region}) with a single $\beta$ model
($S_{0,2}=0$ in equation~[\ref{eq:beta}]) for $106\arcsec<R<260\arcsec$
from the cluster center; the inner radius is just outside the comet
galaxy.  Since we excluded the central region, the core radius for the
single $\beta$ fit cannot be determined. Thus, we fixed $r_{c1}$ at the
best-fit value from a single $\beta$ model fit to the region excluding
S1 while fitting $\beta_1$ and $S_{0,1}$ (table~\ref{tab:beta}).  We fit
the spectrum of the fan-shaped region with one thermal model, and derive
the temperature $T_{\rm out}$, and the density $n_{{\rm gas}, 1, 1}$
(equation~[\ref{eq:rhogas_ob1}]). Using the results and assuming that
the cluster is spherically symmetric for $R>106\arcsec$, we can estimate
the projected spectrum of the outer region of the cluster on the comet
galaxy region.

We fit the spectrum within the 8\farcs9 circle centered on the comet
galaxy (head of the comet in figure~\ref{fig:region}) with two thermal
models; one of them is the spectrum of the outer region of the cluster
projected on the comet galaxy and is fixed.  The deprojected temperature
is quite low, $T = 1.4$ keV, if we assume that the metal abundance is
the same as the ambient gas ($Z=0.47\: Z_\odot$; table~\ref{tab:sp}).
This temperature is more consistent with the gas in a large galaxy or
small group, rather than a cluster.

\begin{figure}
\FigureFile(80mm,80mm){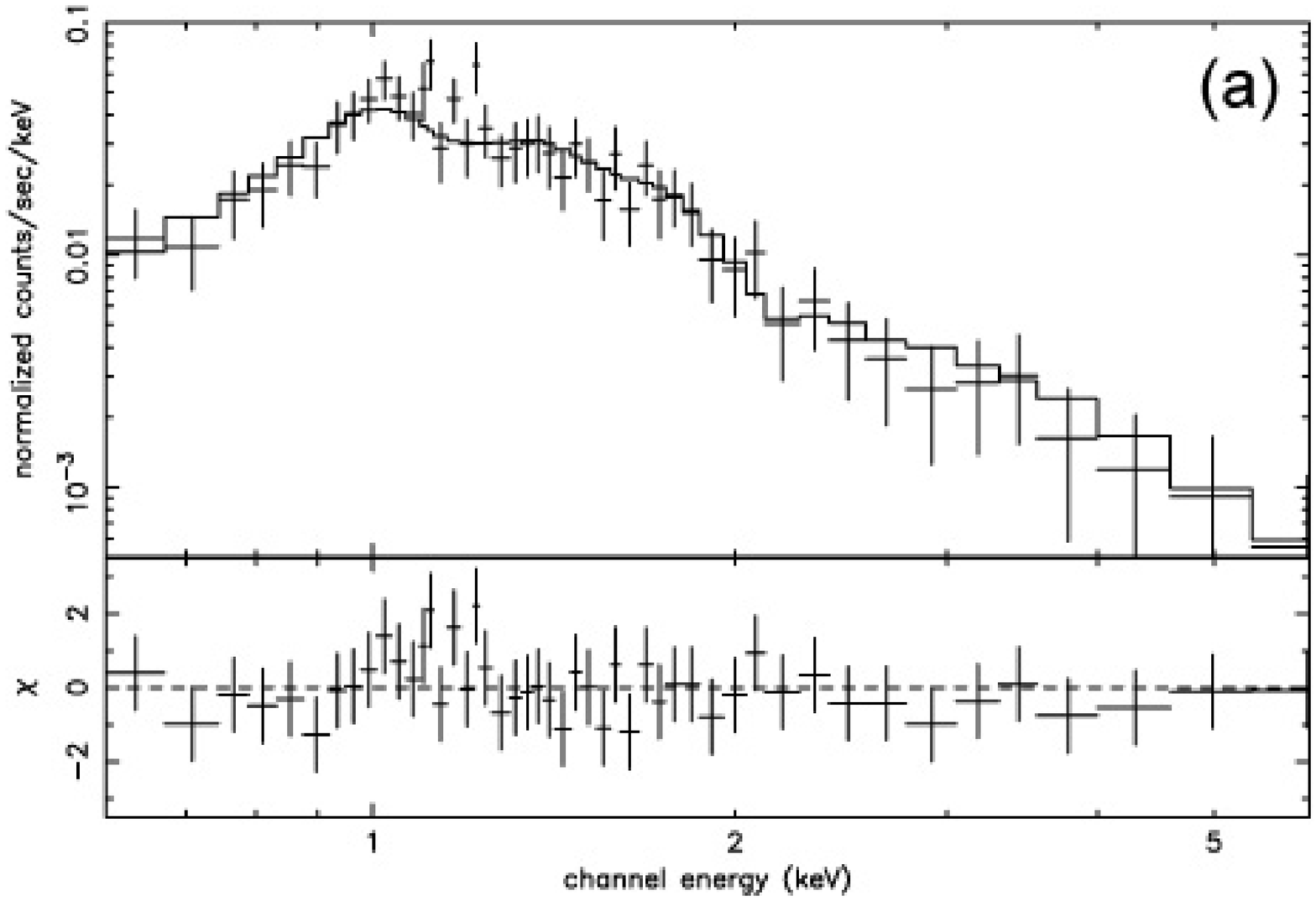} \FigureFile(80mm,80mm){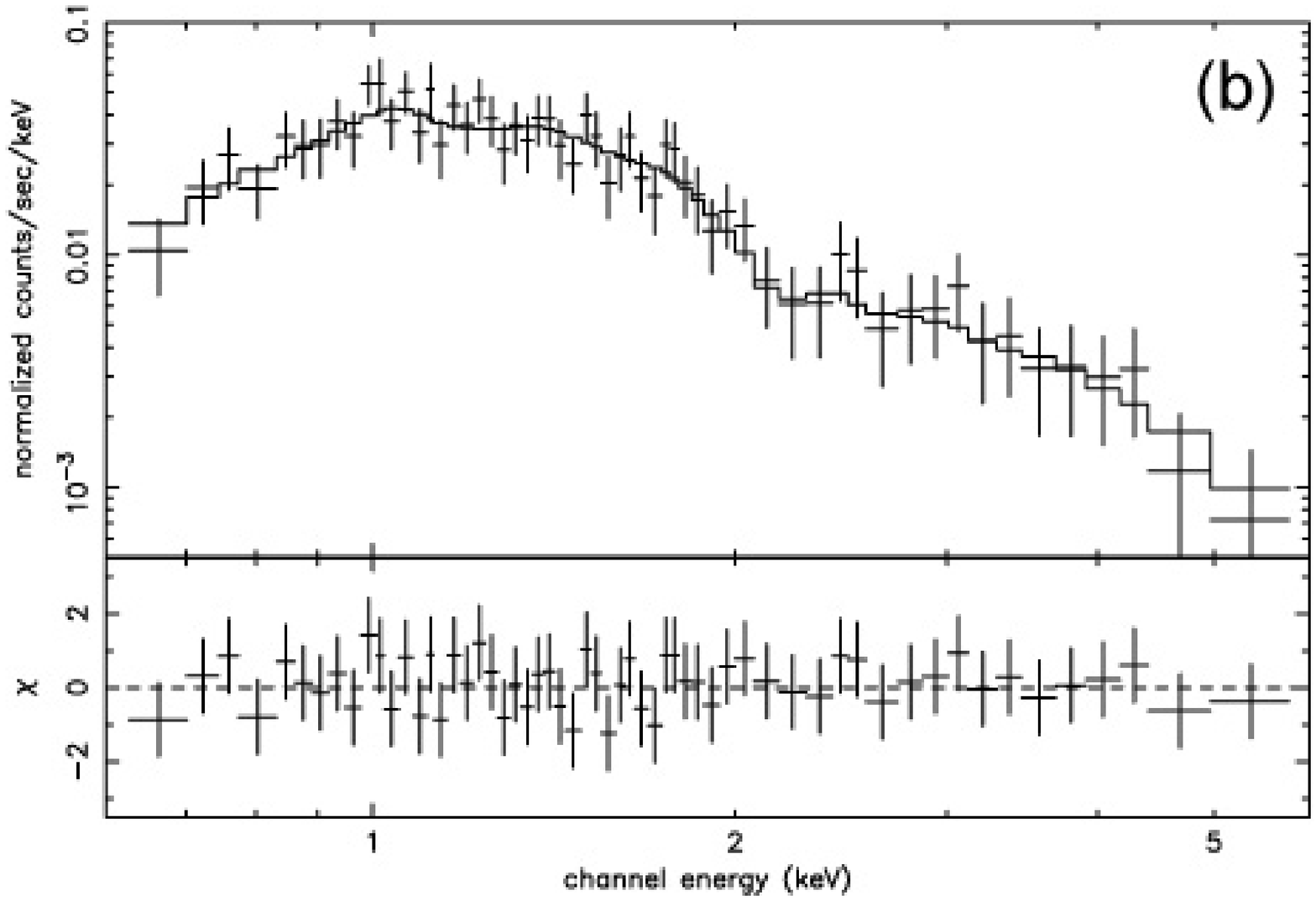}
\caption{(a) Upper panel shows the X-ray spectral data and best-fit
model for the tail region (the solid rectangle in
figure~\ref{fig:region}), while the lower panel plots the residuals
divided by the $1\sigma$ errors. (b) Same as (a), but the spectrum is
for the surrounding $74''\times 49''$ rectangle (the dotted rectangle in
figure~\ref{fig:region}) excluding the tail region.\label{fig:tail_sp}}
\end{figure}

For the tail of the comet galaxy, we fit the spectrum of a $49''\times
22''$ rectangular region shown in figure~\ref{fig:region} (the solid
rectangle) with two thermal models. In order to consider the projection
of the spectra of hotter gas surrounding the tail, we fixed the first of
the two temperatures at the temperature of a surrounding $74''\times
49''$ rectangle (the dotted rectangle in figure~\ref{fig:region})
excluding the tail ($T_{\rm out}$). We also fixed the normalization of
the hotter component at the value expected from the normalization of the
spectrum of the $74''\times 49''$ rectangle. If we assume that the metal
abundance of the tail is same as that of the ambient gas ($Z=0.47\:
Z_\odot$), the deprojected temperature of the tail is $\sim 1.5$~keV
(table~\ref{tab:sp}), which is very similar to that of the galaxy at the
head.  If the gas were simply pulled from the galaxy, one might have
expected that adiabatic expansion would have reduced the temperature
further. The spectra of both regions are shown in
figure~\ref{fig:tail_sp}.

Since the gas in the head and tail of the comet galaxy may come from the
stars in the galaxy, its metal abundance may be higher than the average
value of the cluster. In table~\ref{tab:sp}, we also show the results of
spectral fits obtained on the assumption that the metal abundance of the
gas in the head and tail is $Z=1\: Z_\odot$ and that of the surrounding
gas is $Z=0.47\: Z_\odot$ (the cluster average). Within the errors, the
results are not different from those derived on the assumption that
$Z=0.47\: Z_\odot$ for the head and tail.

\section{Discussion}

\subsection{Kinematics of the Comet Galaxy}
\label{sec:com_kin}

\begin{figure}
\begin{center}
    \FigureFile(80mm,80mm){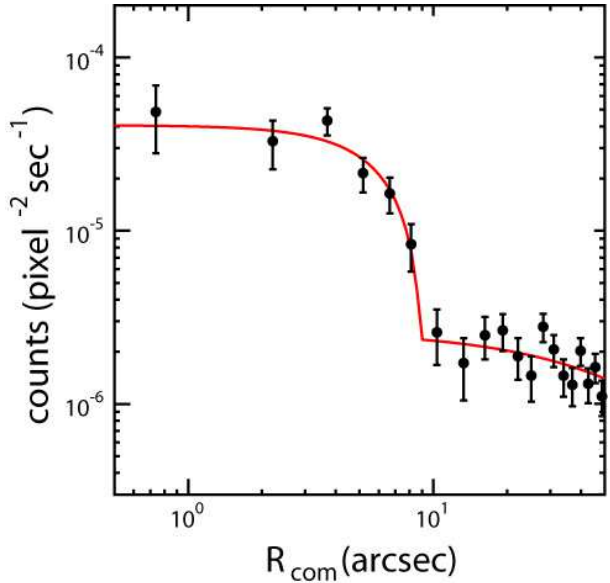} 
\end{center}
\caption{Surface brightness as a function of the distance from the
center of the comet galaxy for 0.3--10~keV for the 90\arcdeg~sector
shown as S2 in figure~\ref{fig:region};
$R_{\rm com}=0$ corresponds to the apex of the S2 region. Error bars are
$1\sigma$ Poisson uncertainties.  The model fit is shown as a solid line
(see text). \label{fig:front}}
\end{figure}

Figure~\ref{fig:front} shows the surface brightness profile for the
region in front of the comet galaxy (S2 in figure~\ref{fig:region}).
The distance, $R_{\rm com}$, is measured from the center of the X-ray
surface brightness contour at a radius of $\sim 8\arcsec$ from the
galaxy center. The selected center is $\sim 2\arcsec$ away from the
X-ray peak because of the asymmetry of X-ray emission.  At $R_{\rm
com}=9\arcsec$, there is a sharp discontinuity in the X-ray surface
brightness.  The hardness ratio map (figure~\ref{fig:HR}) suggests that
there is a corresponding temperature jump (see $T_{\rm out}$ and $T_{\rm
in}$ shown below), with the denser gas being cooler.  The spectral fit
to the comet galaxy head emission is consistent with the dense gas being
quite cool.  Since the brighter (higher density) region has a cooler
temperature, the discontinuity appears to be a cold front, which are
often observed in clusters \citep{mar00,vik01b}. In this subsection, we
estimate the velocity of the comet galaxy from the stagnation condition
at the cold front.

\subsubsection{A Simple Case}
\label{sec:simple}

For the sake of simplicity, we assume that the actual distance from the
cluster center to the comet galaxy ($r_{\rm gal}$) is the same as the
projected distance ($b_{\rm gal}=143$~kpc), which is the lower boundary
of the possible range ($r_{\rm gal}\geq b_{\rm gal}=143$~kpc).  The ICM
density profile can be derived from the result of the single $\beta$
model fit for the fan-shaped $106\arcsec<R<260\arcsec$ region in
section~\ref{sec:comet} (the second line of table~\ref{tab:beta}),
\begin{equation}
\label{eq:n_e}
 n_{\rm gas}(r) = n_{\rm gas,1,1}[1+(r/r_{c1})^2]^{-3 \beta_1/2}\:,
\end{equation}
where $n_{\rm gas,1,1}=9.8_{-1.1}^{+1.1}\times 10^{-2}\rm\:
cm^{-3}$. The result of the fit is shown in figure~\ref{fig:front} (the
solid line for $R_{\rm com}>9\arcsec$).  The density ahead of the galaxy
is $n_{\rm out}=2.0_{-0.2}^{+0.2}\times 10^{-3}\rm\: cm^{-3}$. For the
temperature outside the galaxy, we use the temperature of the fan-shaped
region, which is $T_{\rm out}=4.7_{-0.8}^{+1.1}$~keV.  The stagnation
condition relates the pressure $P_{\rm out}$ far in front of the cold
front with the pressure $P_{s}$ at the stagnation point at the leading
edge of the cold front.  Because of gradients in the overall cluster
density, it is difficult to determine $P_{\rm out}$ exactly, but it is
likely to be smaller than the value determined from the product of
$n_{\rm out}$ and $T_{\rm out}$ immediately in front of the comet
galaxy.

Subtracting the emission around the galaxy, we found that the density
inside the galaxy is almost constant at $n_{\rm
in}=2.8_{-0.2}^{+0.2}\times 10^{-2}\rm\: cm^{-3}$ (the solid line for
$R_{\rm com}<9\arcsec$ in figure~\ref{fig:front}).  The temperature inside
the comet galaxy is $T_{\rm in}=1.4_{-0.1}^{+0.3}$~keV
(table~\ref{tab:sp}).  Thus, the ratio of the stagnation pressure $P_s$
to the pressure far in front is at least $P_{s}/P_{\rm
out}=4.1_{-0.9}^{+1.3}$.  The stagnation condition depends on the Mach
number of the cold front relative to the upstream gas (${\cal M}$) as:
\begin{equation}
\label{eq:vik01b}
\frac{P_{\rm s}}{P_{\rm out}} = \left\{
\begin{array}{ll}
{\cal M}^2\left(\frac{\gamma+1}{2}\right)^{(\gamma+1)/(\gamma-1)}
\left(\gamma-\frac{\gamma-1}{2{\cal M}^2}\right)^{-1/(\gamma-1)} &
\qquad {\cal M}>1 \\
\left(1+\frac{\gamma-1}{2}{\cal M}^2\right)^{\gamma/(\gamma-1)} &
\qquad {\cal M} \le 1 \\
\end{array}
\right.
\end{equation}
where $\gamma$ (=5/3) is the adiabatic index \citep{lan59,vik01b}.  The
resulting Mach number and galaxy velocity are ${\cal M} =
1.6^{+0.3}_{-0.2}$ and $v_{\rm gal}=1.7_{-0.3}^{+0.4}\times 10^3\rm\:
km\: s^{-1}$.  Since the values of $r_{\rm gal}$ and $P_{s}/P_{\rm out}$
we adopted are the lower limits, the values of ${\cal M}$ and $v_{\rm
gal}$ are also the lower limits.

\subsubsection{Ram-Pressure Stripping}

Ram-pressure stripping can have a strong influence on the evolution of
galaxies in clusters \citep{gun72,fuj99,fuj01}.  The morphology of the
comet galaxy suggests that the galaxy is affected by ram-pressure
stripping. A similar galaxy is found in A2125 \citep{wan04}.  If the
comet galaxy is at $r_{\rm gal}=b_{\rm gal}$ (see
section~\ref{sec:simple}), the ram-pressure on the comet galaxy from the
outside ICM ($P_{\rm ram}=\rho_{\rm out}v_{\rm gal}^2$, where $\rho_{\rm
out}$ is the mass density corresponding to $n_{\rm out}$) is $P_{\rm
ram}=1.2_{-0.3}^{+0.5}\times 10^{-10}\rm\: g\: cm^{-1}\: s^{-2}$, based
on the velocity derived above and the cluster gas density profile.

The gas in the comet galaxy would be stripped when
the ram-pressure becomes larger than the static pressure of the galaxy:
\begin{equation}
\label{eq:ram}
 P_{\rm ram}>P_{\rm gal}\equiv \rho_{\rm in}
\frac{k_{\mathrm{B}} T_{\rm in}}{\mu m_{\mathrm{p}}}\;,
\end{equation}
where $\rho_{\rm in}$ is the mass density corresponding to $n_{\rm in}$,
$k_{\mathrm{B}}$ is the Boltzmann constant, $\mu$ (=0.6) is the mean
molecular weight, and $m_{\mathrm{p}}$ is the proton mass
\citep{nul82,mor01a,sar02}.  Since $P_{\rm gal}=1.2_{-0.1}^{+0.2}\times
10^{-10}\rm\: g\: cm^{-1}\: s^{-2}$, the comet galaxy marginally
satisfies the condition~(\ref{eq:ram}).

\subsection{The Origin of the Gas around the Comet Galaxy}

In this subsection, we consider whether the gas around the comet galaxy
can be supplied from the stars in the galaxy. The $B$--band absolute
magnitude of the comet galaxy is $-21.3$\footnote{NED:
http://nedwww.ipac.caltech.edu/}.  Since the stellar mass-loss rate per
unit $B$--band luminosity is given by $\dot{m}_\star\approx1.5\times
10^{-11}\: M_\odot\:{\rm yr^{-1}}\: L_\odot^{-1}$ \citep{fab76}, the
total mass-loss rate of the comet galaxy is $\dot{M}_\star\approx 0.81\:
M_\odot\rm\: yr^{-1}$.  Since the mass of the gas in the comet galaxy
(the head of the comet) is $M_{\rm gal,gas}= 7.2_{-0.4}^{+0.4}\times
10^9\: M_\odot$, the gas would be supplied by the stars over the time
$t_\star=M_{\rm gal,gas}/\dot{M}_\star\approx 8.9\times 10^9$~yr. The
time-scale is marginally smaller than the age of the Universe, and thus
the gas could be supplied by the stars. However, we did not include the
gas stripped from the comet galaxy in this estimate.  Assuming that the
tail of the comet galaxy is roughly a cylinder with a radius of 16~kpc
and the length of 71~kpc (the tail region shown in
figure~\ref{fig:region}), the gas mass of the tail is $1.5\times
10^{10}\: M_\odot$, which is larger than $M_{\rm gal,gas}$.  Since the
temperature of the tail is much smaller than that of the surrounding
region ($\sim 4$~keV; table~\ref{tab:sp}), most of the gas came from the
comet galaxy. This means that most of the gas around the comet galaxy
(head and tail; $\sim 2.2\times 10^{10}\: M_\odot$) did not originate in
the stars in this galaxy at present day loss rates, although there is
uncertainty of the mass loss rate expecially at the early phase of
galaxy evolution.  The comet galaxy probably was the central galaxy of a
small cluster or group, and the gas around it probably originated as
intracluster or intragroup medium.

We note that the above conclusion does not depend on the assumed metal
abundance of the gas in the head and tail of the galaxy
($Z=0.47~Z_\odot$). Even if we assume that the metal abundance of the gas
in the head and tail is higher than the surrounding gas and is $Z=1
Z_\odot$, their gas masses are reduced by only $\sim 15$\%.

We also consider the case where the tail is clumpy. Here, we assume that
the tail is isothermal but consists of two different density gas,
$\rho_1$ and $\rho_2$. The gas with density $\rho_1$ fills the volume $f
V_{\rm tail}$ and that with density $\rho_2$ filles the volume $(1-f)
V_{\rm tail}$, where $V_{\rm tail}$ is the total volume of the tail and
$f$ is the parameter ($0\leq f \leq 1$). If $\rho_1\gg \rho_2$, the
luminosity of the tail is given by $L_{\rm tail}\propto \rho_1^2 f
V_{\rm tail}$. On the other hand, if we assume that the tail is uniform
(the density is $\rho_0$ and the mass is $M_{\rm tail, 0}=\rho_0 V_{\rm
tail}$), the luminosity is represented by $L_{\rm tail}\propto \rho_0^2
V_{\rm tail}$. Thus, for a given $L_{\rm tail}$, the density of the
clumps is written as $\rho_1 = \rho_0 f^{-1/2}$, and the mass of the
clumpy tail is approximately given by $M_1=\rho_1 f V_{\rm tail}=M_0
f^{1/2}$. This means that if the tail is highly clumpy, the tail mass
could be smaller than that derived above ($M_{\rm tail, 0}=1.5\times
10^{10}\: M_\odot$). However, such clumps would easily be destroyed
through hydrodynamical interactions with the surrounding gas
(e.g. \authorcite{mat90} \yearcite{mat90}).

\subsection{The cD Galaxies of A2670 and A2107}
\label{sec:cD}

 From the hardness ratio maps (figure~\ref{fig:HR}) and surface brightness
profiles (figure~\ref{fig:surf}), we know that the cool, dense
interstellar medium (ISM) of A2670 and A2107 is confined within
3\arcsec~and 15\arcsec~from the galaxy centers (X-ray peaks),
respectively. The pressure ratio at the boundary is $P_{\rm ISM}/P_{\rm
ICM}=0.7_{-0.4}^{+0.9}$ for A2670 and $0.7_{-0.2}^{+0.5}$ for A2107.
Thus the pressure is consistent with being continuous at the boundary.
The allowable velocities derived from a stagnation condition
(equation~[\ref{eq:vik01b}]) are $\lesssim 1400\:\rm km\: s^{-1}$ for
A2670 and $\lesssim 1000\:\rm km\: s^{-1}$ for A2107.  Since these
limits are still much larger than the observed radial peculiar
velocities of the cDs in their clusters ($v_{\mathrm{P}}=433\:\rm km\:
s^{-1}$ for A2670, and $270\:\rm km\: s^{-1}$ for A2107;
\authorcite{oeg01} \yearcite{oeg01}), these limits are not very
constraining.

The ISM masses are $M_{\rm ISM}=7.8_{-2.9}^{+2.9}\times 10^8\: M_\odot$
and $5.5_{-0.9}^{+0.9}\times 10^9\: M_\odot$ for A2670 and A2107,
respectively. In this estimation, we assumed that the metal abundance of
the ISM is the same as the cluster average. If we assume that the metal
abundance of the ISM is $Z=1\: Z_\odot$ as in the case of the comet
galaxy, the ISM masses are reduced by only $\lesssim 15$\%, which does
not affect the following discussion. The $B$--band absolute luminosities
of the cD galaxies are $-22.7$ and $-21.6$ for A2670 and A2107,
respectively. As we did for the comet galaxy, we derive the time-scale
for the stars in the cD galaxies to supply the ISM. For A2670,
$t_\star=2.8\times 10^8$~yr, and for A2107, $t_\star=5.1\times
10^9$~yr. The time-scales are smaller than or comparable to the typical
age of a cluster ($\sim 3$--10~Gyr; e.g. \authorcite{kit96}
\yearcite{kit96}).  On the other hand, only a small fraction of the
stellar mass loss would occur within the small volume occupied by the
cooler gas today.  Thus, stellar mass loss would only be adequate if the
gas was ejected before the cD galaxies were located in their present
environment, and the ISM was compressed by the high ICM pressure to its
current small volume.  In any case, the cooling times of the ISM are
$t_{\rm cool}=2.7_{-1.4}^{+2.8}\times 10^8$~yr for A2670, and
$8.3_{-2.2}^{+5.9}\times 10^8$~yr for A2107. Thus, while stars in the cD
can marginally supply the ISM before cooling becomes effective for
A2670, they cannot for A2107.

For A2670, the size of the ISM region is comparable to those of NGC~4874
and 4889 in the Coma cluster ($\sim 3$~kpc; \authorcite{vik01c}
\yearcite{vik01c}). Both A2670 and Coma are merging clusters. Thus, a
compact ISM may be common for merging clusters. For the two galaxies in
Coma, it has been shown that $t_{\rm cool}<t_\star$ and some heating
source is required to prevent radiative cooling \citep{vik01c}.  Since
$t_{\rm cool}\approx t_\star$, the ISM of the cD galaxy in A2670 may
also be affected by the heating source. Although the size of the ISM
region is larger (12~kpc), radiative cooling must be suppressed as well
for A2107 ($t_{\rm cool}<t_\star$). These observations show the
similarity between `cooling flow' clusters and dynamically young merging
clusters in terms of the necessity of some heating sources. For the
dynamically young clusters, active galactic nuclei (AGNs) are not good
candidates for heating sources because AGN jets would deposit their
energy outside the small ISM regions (e.g. \authorcite{omm04}
\yearcite{omm04}). Thus, as discussed by \citet{vik01c}, thermal
conduction may be the heating source. In fact, the time-scales of the
thermal conduction are short; $t_{\rm cond}\approx 8.5\times 10^6$~yr
for A2670, and $1.9\times 10^7$~yr for A2170, which are derived using
equation~(9) in Fujita and Goto (2004). Note that the saturation of
thermal conduction is generally not important for cD galaxies because
the electron density is high around them.  Since $t_{\rm cond}\ll
t_\star$, thermal conduction can supply enough energy to the ISM. On the
other hand, at least for A2107, the ISM structure seen in the surface
brightness map (figure~\ref{fig:cen}b) and that seen in the hardness ratio
map (figure~\ref{fig:HR}b) are different. 
This indicates that the ISM is not in pressure equilibrium, which may be
due to the dynamical motion around the galaxy. Cluster mergers may
induce such motion, which may finally lead to turbulence and heat the
cluster core through the turbulent mixing as is shown in the `tsunami'
model \citep{fuj04c,fuj05a}.

\section{Conclusions}

We present an analysis of Chandra observations of the galaxy clusters
A2670 and A2107. The peculiar velocities of the cD galaxies are
unusually large ($>200~\rm\: km\: s^{-1}$) and the clusters are
undergoing mergers.  To the west of the cD galaxy of A2670, we find a
galaxy having a comet-like X-ray structure (comet galaxy). The leading
edge of the structure is a cold front. From the pressure profile across
the cold front, we estimate the velocity of the galaxy in the cluster
A2670. The mass of X-ray gas in the comet-like structure is too large to
have been produced by stellar mass loss within the galaxy itself.  Thus,
it is likely that the galaxy was at the center of a small cluster or
group, and that its intracluster or intragroup medium is being stripped
by ram-pressure.  The sizes of the cooler, X-ray interstellar medium
(ISM) regions of the cD galaxies in A2670 and A2107 are very small.
This is similar to the brightest galaxies in the Coma cluster, which is
also a merging cluster.  Since the cooling time of the ISM is small,
there must be some heating sources in the galaxies.  The compactness of
the ISM indicates that the heating sources are not the AGNs in the
galaxies.  We suggest thermal conduction or hydrodynamical heating by
`tsunamis' as possible heat sources to balance the observed X-ray
emission.

\vspace{5mm}

We are grateful to Tracy Clarke for useful comments.  Y. F. is supported
in part by a Grant-in-Aid from the Ministry of Education, Culture,
Sports, Science, and Technology of Japan (14740175).  Support for this
work was provided by the National Aeronautics and Space Administration
primarily through Chandra award GO4-5137X, but also through
GO4-5133X, GO4-5149X and GO4-5150X, issued by the Chandra X-ray
Observatory, which is operated by the Smithsonian Astrophysical
Observatory for and on behalf of NASA under contract NAS8-39073.
G. R. S. acknowledges the receipt of an ARCS fellowship.


\end{document}